\PassOptionsToPackage{numbers,sort&compress}{natbib}
\documentclass[12pt]{elsarticle}
\usepackage[margin=2.5cm]{geometry}
\usepackage{enumitem}
\newlist{inlinelist}{enumerate*}{1}
\setlist*[inlinelist,1]{%
  label=(\roman*),
}
\usepackage{dirtytalk}
\usepackage{soul}
\usepackage{hhline}
\usepackage{multirow}
\usepackage{xcolor}
\usepackage{graphicx}
\usepackage{upgreek}
\usepackage{subcaption}
\usepackage{amssymb}
\usepackage{amsfonts,amsthm,bm,amsmath} % Math packages
\usepackage{appendix}
\usepackage{times}
\usepackage{caption}

\usepackage{multirow}
\usepackage{xcolor}
\usepackage[linesnumbered,ruled,vlined]{algorithm2e}
\usepackage{tablefootnote}
\usepackage{comment}
\usepackage{bm}
\usepackage{tabularx}
\usepackage{caption}
\usepackage{array}
\usepackage{dsfont}

\SetCommentSty{mycommfont}

\SetKwInput{KwInput}{Input}                % Set the Input
\SetKwInput{KwOutput}{Output}              % set the Output

\usepackage[pagewise]{lineno}

\usepackage[colorlinks]{hyperref} 
\hypersetup{ 
    colorlinks=true,       % false: boxed links; true: colored links
    linkcolor=red,          % color of internal links
    citecolor=blue,        % color of links to bibliography
    filecolor=magenta,      % color of file links
    urlcolor=cyan           % color of external links
}

\newcommand{\fref}[1]{Fig.~\ref{#1}}

\newcommand{\eref}[1]{Eq.~(\ref{#1})}

\newcommand{\tref}[1]{Table~\ref{#1}}
\setcitestyle{square}

\usepackage{titlesec}
\usepackage[capitalise]{cleveref}

\titleformat{\paragraph}[hang]{\normalfont\itshape}{\theparagraph}{1em}{}
\titlespacing*{\paragraph}{0pt}{1ex plus 0.1ex minus 0.1ex}{0.5ex plus 0.1ex minus 0.1ex}

\makeatletter
\def\ps@pprintTitle{%
 \let\@oddhead\@empty
 \let\@evenhead\@empty
 \def\@oddfoot{\reset@font\hfil\thepage\hfil}%
 \let\@evenfoot\@oddfoot}
\makeatother

\begin{document}
\begin{frontmatter}

%Residual-Conditioned Video Diffusion for Spatio-Temporal PDE Surrogates with S-DeepONet Priors
%Two-Stage Residual Learning with S-DeepONet Priors and Video Diffusion for Dynamic PDEs
%From Priors to Residuals: Spatio-Temporal PDE Modeling via S-DeepONet and Conditional Video Diffusion
%Residual Video Diffusion on S-DeepONet Priors for High-Fidelity Spatio-Temporal PDE Prediction
%Hybrid Residual Video Diffusion atop S-DeepONet for High-Fidelity Spatio-Temporal PDE Prediction
\title{Bridging Sequential Deep Operator Network and Video Diffusion: Residual Refinement of Spatio-Temporal PDE Solutions}
\author[]{Jaewan Park$^{1,2}$}
\author[]{Farid Ahmed$^{1,3}$}
\author[]{Kazuma Kobayashi$^{1,3}$}
\author[]{Seid Koric$^{1,2}$
\corref{mycorrespondingauthor}}
\cortext[mycorrespondingauthor]{Corresponding author}
\ead{koric@illinois.edu}
\author[]{Syed Bahauddin Alam$^{1,3}$}
\author[]{Iwona Jasiuk$^{2}$}
\author[]{Diab Abueidda$^{1,4}$}

%\cortext[mycorrespondingauthor2]{Corresponding author}
%\corref{mycorrespondingauthor}
%\ead{abueidd2@illinois.edu}
\address{$^1$ National Center for Supercomputing Applications, University of Illinois at Urbana-Champaign, Urbana, IL, USA \\
$^2$ The Grainger College of Engineering, Department of Mechanical Science and Engineering, University of Illinois at Urbana-Champaign, Urbana, IL, USA \\
$^3$ The Grainger College of Engineering, Department of Nuclear, Plasma \& Radiological Engineering, University of Illinois at Urbana-Champaign, Urbana, IL, USA \\
$^4$ Civil and Urban Engineering Department, New York University Abu Dhabi, United Arab Emirates \\
}

\begin{abstract}

Video-diffusion models have recently set the standard in video generation, inpainting, and domain translation thanks to their training stability and high perceptual fidelity. Building on these strengths, we adopt and train conditional video diffusion as a surrogate for spatio-temporal solution fields governed by partial differential equations (PDEs). Our two-stage surrogate first applies a Sequential Deep Operator Network (S-DeepONet) to produce a coarse, physics-consistent prior from the prescribed boundary or loading conditions. The prior is then passed, together with the original input function, to a conditional video diffusion model that learns only the residual: the point-wise difference between the ground truth and the S-DeepONet prediction. By shifting the learning burden from the full solution to its much smaller residual space, this innovative approach enables diffusion to focus on sharpening high-frequency solution structures without sacrificing global coherence.

The framework is assessed on two disparate benchmarks: (i) vortex-dominated lid-driven cavity flow and (ii) tensile plastic deformation of dog-bone specimens. Across these data sets the hybrid surrogate consistently outperforms its single-stage counterpart, cutting the mean relative $L_2$ error from 4.57\% to 0.83\% for the flow problem and from 4.42\% to 2.94\% for plasticity, a relative improvements of 81.8\% and 33.5\% respectively. The hybrid approach not only lowers quantitative errors but also improves visual quality, visibly recovering sharp transients and fine spatial details.

These novel results show that (i) conditioning diffusion on a physics-aware prior enables faithful reconstruction of localized, high-gradient features, (ii) residual learning simplifies the target distribution, accelerating convergence and enhancing accuracy, and (iii) the same architecture transfers seamlessly from incompressible turbulent flow to nonlinear elasto-plasticity without any problem-specific architectural modifications, highlighting its broad applicability to nonlinear, time-dependent phenomena in engineering and science. 

\end{abstract}


\end{frontmatter}

%\linenumbers

\section{Introduction}
\label{sec:intro}

Neural operator networks have emerged over the past five years as a powerful class of function-to-function surrogates: given boundary data, material coefficients, load histories or previous solution fields, they learn the entire solution operator and return full spatio-temporal fields in a single forward pass \cite{kovachki2023neural}. Due to their nature to retrieve full solution fields, neural operator networks are widely used as a powerful surrogate paradigm for computationally heavy numerical simulations \cite{brunton2020machine, eghbalian2023physics}. One type of neural operator network is the Fourier Neural Operator (FNO) \cite{li2021fourier} which learns global integral kernels in Fourier space. They have been reported to gain three orders‑of‑magnitude speed‑ups on Burgers, Darcy, and incompressible Navier–Stokes benchmarks while retaining high accuracy, even for turbulent super‑resolution. The other type of neural operator is the Deep Operator Network (DeepONet) \cite{lu2021learning}, which has gained popularity due to its generality. It uses a branch–trunk decomposition backed by the universal approximation theorem, which guarantees to learn nonlinear operators from modest data. The original work of DeepONet consists of fully connected neural networks in its' branch and trunk network. Since its debut, a rich ecosystem of problem-aware variants has appeared, experimenting different types of neural network architectures in branch and trunk according to the type of problem it tries to solve. ResUNet-DeepONet is one example, which exploits residual U-Net (ResUNet) in trunk \cite{he2023novel}. Authors have various 2D geometry with load applied at one end, predicted the von Mises stress solution field. Another variant of this deep operator network is the Sequential DeepONet (S-DeepONet) which utilizes recurrent neural network in the branch network \cite{he2024sequential, he2024predictions}. Taking advantage of recurrent neural networks, it handles temporal input functions effectively to construct solution fields in elasto-plastic problems and transient heat‑transfer tasks. These DeepONet variants were also tested with multiphysics setting. With both displacements and heat flux applied simultaneously in one specimen, DeepONet variants successfully reconstructed von Mises stress field and temperature field \cite{kushwaha2024advanced}. Such operator networks pave the way for mesh-free, once-for-all surrogates that evaluate in milliseconds while comprehending complex boundary conditions.

Generative AI has advanced through a sequence of paradigm shifts, each widening the scope of what machines can synthesize. Early breakthroughs such as the Variational Auto-Encoder (VAE) and its conditional extension, the CVAE \cite{sohn2015learning, kingma2013auto}, introduced latent-variable modeling capable of conditional image and sequence generation. Shortly thereafter, Generative Adversarial Networks (GANs) \cite{goodfellow2014generative} reframed generation as a minimax game between two networks, yielding photo-realistic imagery and inspiring hundreds of domain-specific variants. The quest for discrete, information-rich latents produced the VQ-VAE \cite{van2017neural}, whose quantized codebook enabled faithful reconstruction of images, speech, and even short video clips. The field pivoted again with Denoising Diffusion Models \cite{sohl2015deep, ho2020denoising, ho2022cascaded, ho2022classifier, dhariwal2021diffusion}, which interpret generation as time-reversed diffusion and achieve state-of-the-art sample quality while providing an explicit likelihood. Progress accelerated when the forward-inverse dynamics were re-formulated as a stochastic differential equation (SDE) \cite{song2021score}, unifying score matching and denoising objectives and paving the way for subsequent refinements. Most notably, Karras et al. systematically optimizes noise schedules and further streamlines sampling through a Heun-style second order solver eschewing the Euler predictor used in earlier works by introducing their own framework, Elucidated Diffusion models (EDM) \cite{karras2022elucidating}. Diffusion foundations have since scaled from still images to coherent spatio-temporal synthesis: Video Diffusion Models integrate joint image–video training and conditional sampling to produce the first large-scale text-to-video generations with smooth dynamics \cite{voleti2022mcvd, ho2022video, xing2024survey}, while Imagen Video cascades base and super-resolution diffusers to push video resolution and photorealism even further \cite{ho2022imagenvideo}. In parallel, the Schrödinger Bridge perspective recasts diffusion as an entropy-regularized optimal-transport problem, unifying score matching, flow matching, and bridge dynamics under a stochastic control lens that naturally extends to physical time-series data \cite{de2021diffusion}. Today, latent-variable VAEs, adversarial GANs, discrete VQ-VAEs, probabilistic diffusion, and optimal-transport bridges constitute a rich generative toolkit, positioning AI to model not only images and language but also the complex spatio-temporal fields encountered in scientific and engineering simulations.

While neural operators such as FNO and DeepONet excel at capturing the global structure of parametric PDE solutions, their spectral bias tends to smear sharp gradients and under‐predict fine‐scale energy \cite{rahaman2019spectral}. Recent studies therefore augments a deterministic operator (or low‑resolution solver) with a stochastic generative corrector that recovers the missing high‑frequency content. Early examples include the physics‑guided diffusion model \cite{lu2024generative} which learns to super‑resolve coarse PDE outputs and then nudges the samples toward a low residual with respect to the governing equations, and NVIDIA’s two‑stage CorrDiff framework \cite{mardani2025residual}, where a regression UNet supplies a mean field that a conditional diffusion model refines into kilometer‑scale weather realizations. LatentPINN \cite{taufik2025latentpinns} uses a latent diffusion prior to draw PDE coefficients that feed a physics‑informed neural network, enabling one PINN to generalize across parameter families without retraining, whereas the Generative Adversarial Neural Operator (GANO) \cite{rahman2022generative} casts the generator–discriminator game directly in function space to sample stochastic solutions consistent with learned operator mappings. Variational auto‑encoder variants also appear. The physics‑informed DVAE from Glyn‑Davies et al. \cite{glyn2024varphi} assimilates unstructured observations by coupling a dynamical VAE with latent PDE priors. Furthermore, the Physics‑Informed Diffusion Models (PIDM) framework \cite{bastek2024physics} injects the PDE residual directly into the denoising score‑matching objective, cutting physics violations by up to two orders of magnitude without post‑processing. In turbulence modeling, Oommen et al. demonstrate that conditioning a diffusion corrector on neural operator outputs markedly improves the predicted energy spectra and stabilizes long autoregressive roll‑outs \cite{oommen2024integrating}, using FNO, UNet, and Time-Conditioned UNet (TC-UNet) \cite{ovadia2023real} based operator networks. 

Collectively these efforts confirm the promise of marrying deterministic operators with probabilistic refiners, yet current pipelines often treat the two stages as loosely coupled post‑processors or require the generative model to learn the entire solution field from scratch. Moreover, existing studies remain confined to 2D, image-based diffusion architectures; to the best of our knowledge, no work has yet coupled neural operator outputs with a video-diffusion model capable of synthesizing full space–time fields. A tighter, residual‑focused integration framework that lets the operator provide coarse coherence while the generative video decoder concentrates exclusively on the small but crucial error manifold remains unexplored and motivates the present study. Inspired by the operator‑conditioned paradigm of \cite{oommen2024integrating}, we address these gaps by: (i) training the diffusion stage only on the residual, (ii) conditioning on a spatio-temporal S-DeepONet prior rather than slice-wise forecasts, and (iii) upgrading the corrector to a video-diffusion model that synthesizes full space–time volumes.

This paper presents an unexplored hybrid surrogate framework that couples a Sequential Deep Operator Network (S-DeepONet) with a conditional video diffusion decoder for data-driven prediction of spatio-temporal fields governed by nonlinear partial differential equations. The operator network maps the boundary measurements to a coarse physics-consistent prior video, whereas the diffusion stage iteratively refines this scaffold in pixel space, recovering the high-frequency details and sharp transients that the operator network alone cannot resolve. We detail the architecture, conditioning pathway, and training protocol, and we test the method on two disparate benchmarks (i) laminar to turbulent lid-driven cavity flow and (ii) elasto-plastic deformation of dogbone specimens, thereby demonstrating seamless applicability across nonlinear fluid and solid mechanics. Systematic ablations against stand-alone S-DeepONet and standalone diffusion reveal that each component excels at complementary aspects of the task, yet their integration delivers both the lowest quantitative errors and the highest visual fidelity. Together, these results position the proposed framework as a practical, physics-aware route to high-accuracy field prediction, effectively combining two models with synergy.

\section{Methods}
\label{sec:methods}

\subsection{Training data generation}
\label{sec:training_data_generation}

\subsubsection{Lid driven cavity flow}
\label{sec:ldc_training_data_generation}

\begin{figure}[h!]
    \centering
    \begin{minipage}[b]{0.4\textwidth}
        \centering
        \includegraphics[width=\textwidth]{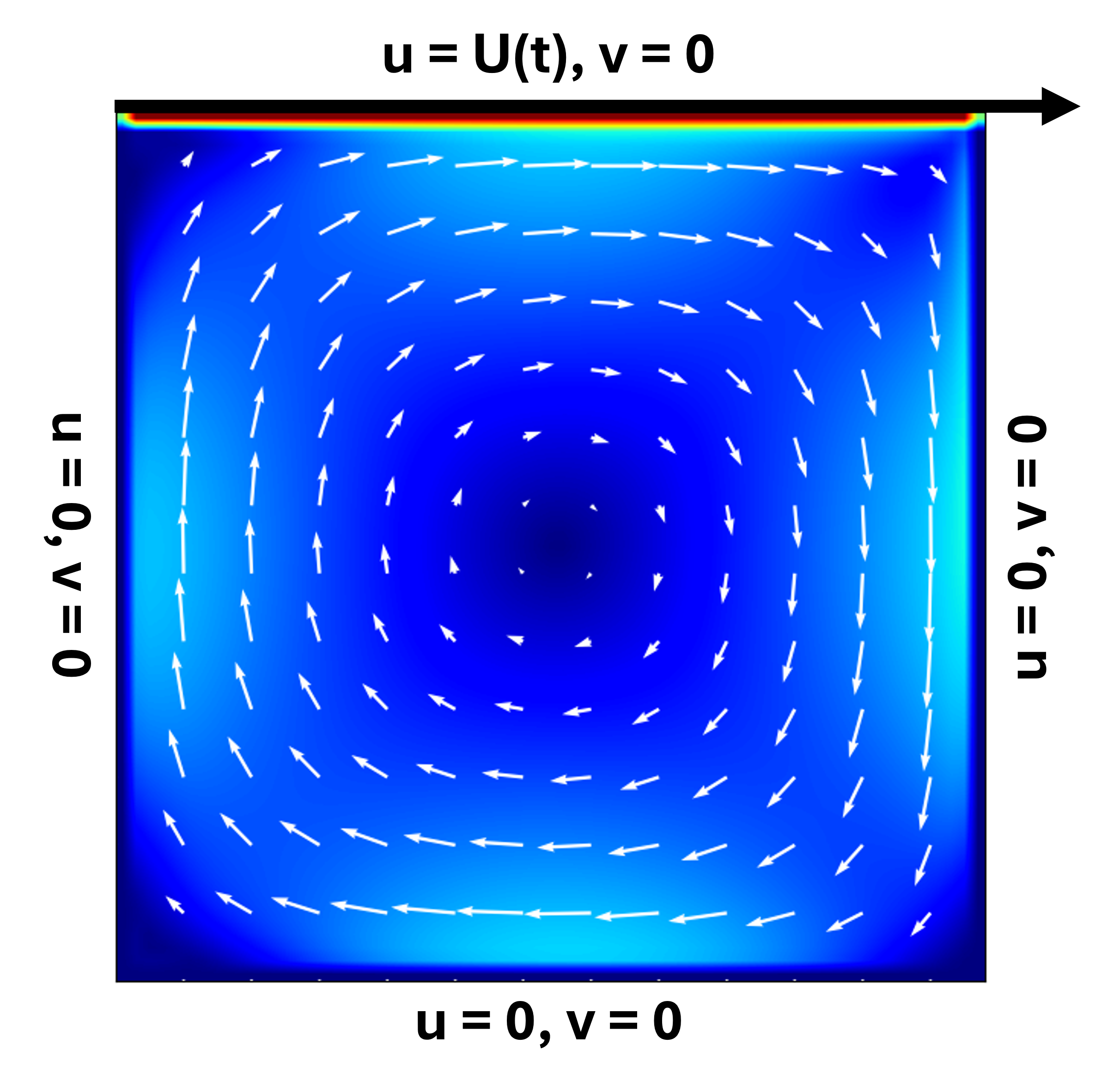}
        \subcaption{Lid driven cavity flow domain example}
        \label{fig:ldc_sample}
    \end{minipage}
    \hfill
    \begin{minipage}[b]{0.5\textwidth}
        \centering
        \includegraphics[width=\textwidth]{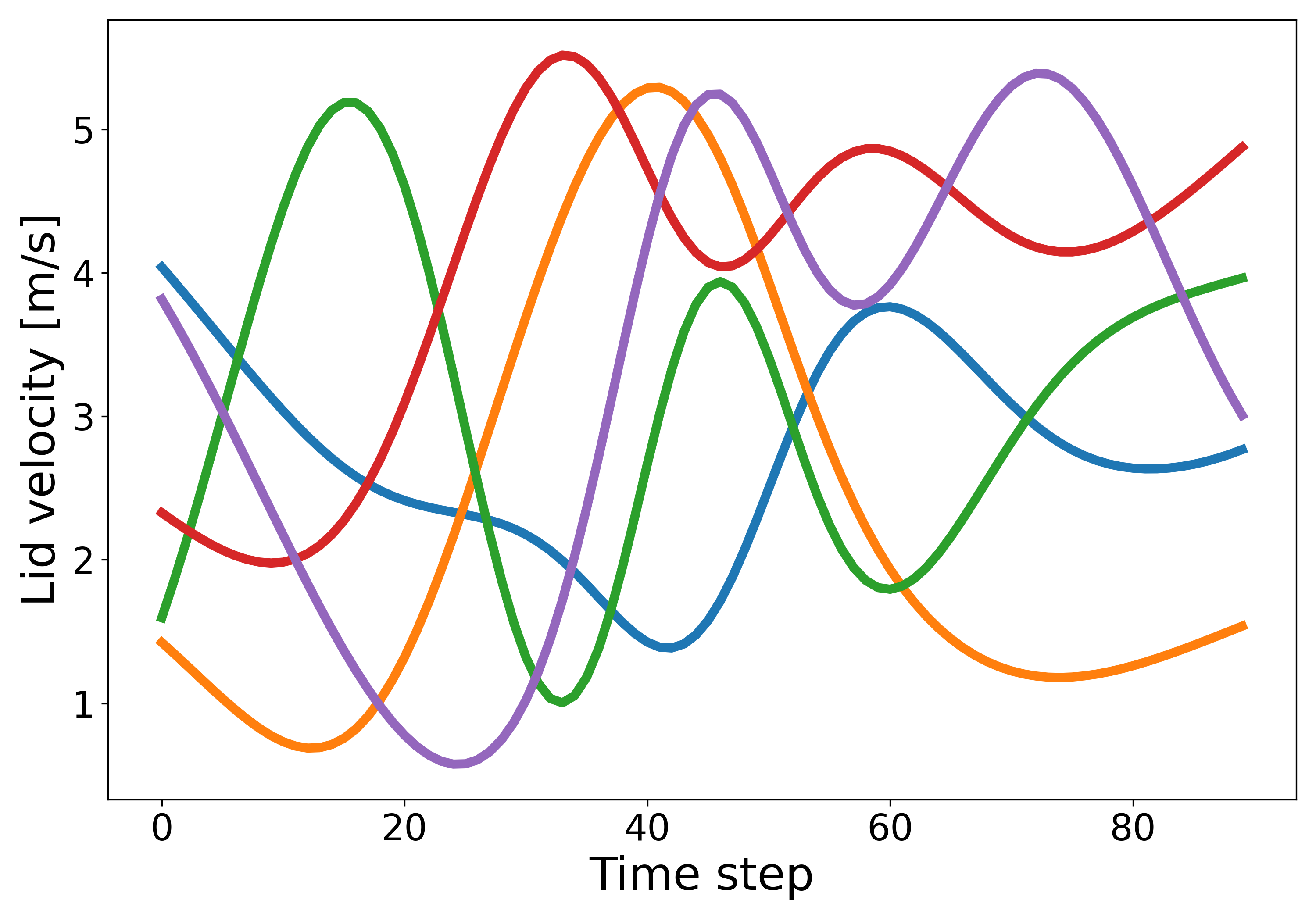}
        \subcaption{Five examples of lid velocity.}
        \label{fig:lid_vel}
    \end{minipage}
    \caption{Geometry and representative lid velocity of fluid dynamics simulation.}
    \label{fig:ldc_sample_lid_vel}
\end{figure}

%% Modify the description for the Reynolds-averaged Navier-Stokes (RANS) model
\begin{comment}
The first training data comprises lid driven cavity flow including turbulence. A square cavity of unit length 0.065m, as shown in \fref{fig:ldc_sample},is solved in two dimensions with the top wall translating rightwards while the other three walls remain stationary. We use the unsteady incompressible Navier–Stokes solver in Ansys-Fluent \cite{AnsysFluent2024} to obtain ground truth videos. The governing equation consists:
\begin{equation}
\begin{aligned}
    \frac{\partial \mathbf{V}}{\partial t}+(\mathbf{V} \cdot \nabla) \mathbf{V}& - \frac{1}{\operatorname{Re}} \nabla^2 \mathbf{V}+\nabla \mathbf{p}=0 \\
    \nabla \cdot \mathbf{V}& = 0
\end{aligned}
\end{equation}

where $\mathbf{V}=f(u, v)$ and the pressure drop $\nabla \mathbf{p}$ is due to the shear driven flow.

Having velocity vector fields as an outcome of the simulation, we calculate stream function for every video in each frame:
\begin{equation}
    u=\frac{\partial \psi}{\partial y}, \quad v=-\frac{\partial \psi}{\partial x}.
\end{equation}
Dirichlet conditions are imposed by setting $\psi = 0$ along all walls. In total 4800 distinct lid velocity histories are simulated, yielding 4800 stream function videos that serve as paired inputs for S-DeepONet (initial guess for the full field) and the video-diffusion corrector (more accurate full field) that will be detailed further in the later sections.
\end{comment}

The first training dataset is generated from a turbulent lid-driven cavity flow simulation. A square cavity of side length 0.065 m, as shown in \fref{fig:ldc_sample}, is modeled in two dimensions, with the top wall translating to the right while the remaining three walls are stationary.

We define the velocity field $\mathbf{u} = (u, v)$ and pressure field $p$, which satisfy the incompressibility condition $\nabla \cdot \mathbf{u} = 0$. The flow dynamics are governed by the incompressible Reynolds-averaged Navier–Stokes (RANS) equations~\cite{rans}, given by:
\begin{equation}
    \frac{\partial \mathbf{u}}{\partial t} + (\mathbf{u} \cdot \nabla)\mathbf{u} = -\nabla p + \nabla \cdot \left[ \left( \nu + \nu_t \right) \left( \nabla \mathbf{u} + \nabla \mathbf{u}^\top \right) \right],
\end{equation}

where $\nu$ is the kinematic viscosity, and $\nu_t$ denotes the eddy viscosity introduced by the turbulence closure model. The effects of turbulence are further characterized using the turbulent kinetic energy (TKE) field $k(x, y, t)$, governed by the following transport equation:

\begin{equation}
\frac{\partial k}{\partial t} + \mathbf{u} \cdot \nabla k = P_k - \varepsilon + \nabla \cdot \left[ \left( \nu + \frac{\nu_t}{\sigma_k} \right) \nabla k \right],
\end{equation}

where $P_k$ is the turbulence production term, $\varepsilon$ is the dissipation rate, and $\sigma_k$ is a model constant.

To induce spatiotemporal variability in the flow field, a time-dependent velocity profile is applied at the top lid. Specifically, the lid velocity varies across simulations according to smoothly changing temporal functions, as illustrated in \fref{fig:lid_vel}, while the remaining boundaries are fixed with no-slip conditions. This dynamic boundary forcing introduces transient unsteady structures and enhances the diversity of the resulting flow patterns.

Using this setup, 4800 simulations were performed, each driven by a unique lid velocity history. For each simulation, the full-field velocity vectors $\mathbf{u}(x, y, t)$ are recorded over time. To obtain a scalar representation of the flow, the stream function $\psi(x, y, t)$ is computed at each time step using:

\begin{equation}
u = \frac{\partial \psi}{\partial y}, \quad v = -\frac{\partial \psi}{\partial x},
\end{equation}
with Dirichlet boundary conditions $\psi = 0$ enforced along all walls. The resulting 4800 stream function sequences constitute the core of the training dataset. Each sequence is paired with its corresponding lid velocity profile and is used to train the S-DeepONet model, which provides an initial full-field estimate. A video-diffusion corrector is subsequently applied to refine the prediction. Details of the model architecture and training procedures are presented in later sections.

\subsubsection{Plastic deformation with time dependent load}
\label{sec:dogbone_training_data_generation}
\begin{comment}
\begin{figure}[h!]
    \centering
    \begin{subfigure}[c]{0.58\textwidth}
        \includegraphics[width=\textwidth]{dogbone_sample.png}
        \caption{Dogbone specimen used in the experiment.}
        \label{fig:dogbone_sample}
    \end{subfigure}
    \hfill
    \centering
    \begin{subfigure}[c]{0.4\textwidth}
        \includegraphics[width=\textwidth]{amp_plot.png}
        \caption{Five examples of boundary displacement conditions.}
        \label{fig:dogbone_load}
    \end{subfigure}
    \caption{Specimen geometry and representative applied boundary conditions.}
    \label{fig:dogbone_sample_load}
\end{figure}
\end{comment}

\begin{figure}[h!]
    \centering
    \begin{minipage}[c]{0.58\textwidth}
        \centering
        \includegraphics[width=\textwidth]{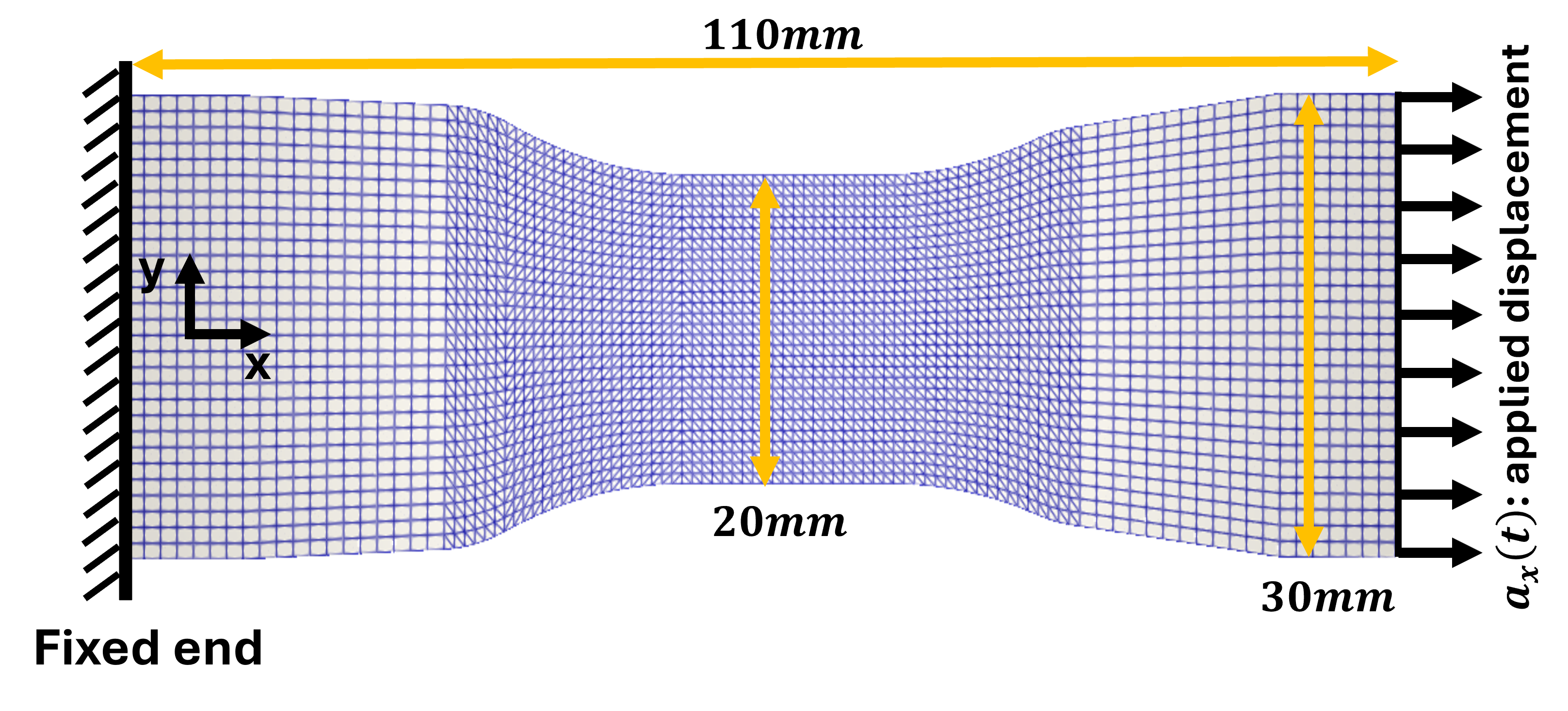}
        \subcaption{Dogbone specimen used in the experiment.}
        \label{fig:dogbone_sample}
    \end{minipage}
    \hfill
    \begin{minipage}[c]{0.4\textwidth}
        \centering
        \includegraphics[width=\textwidth]{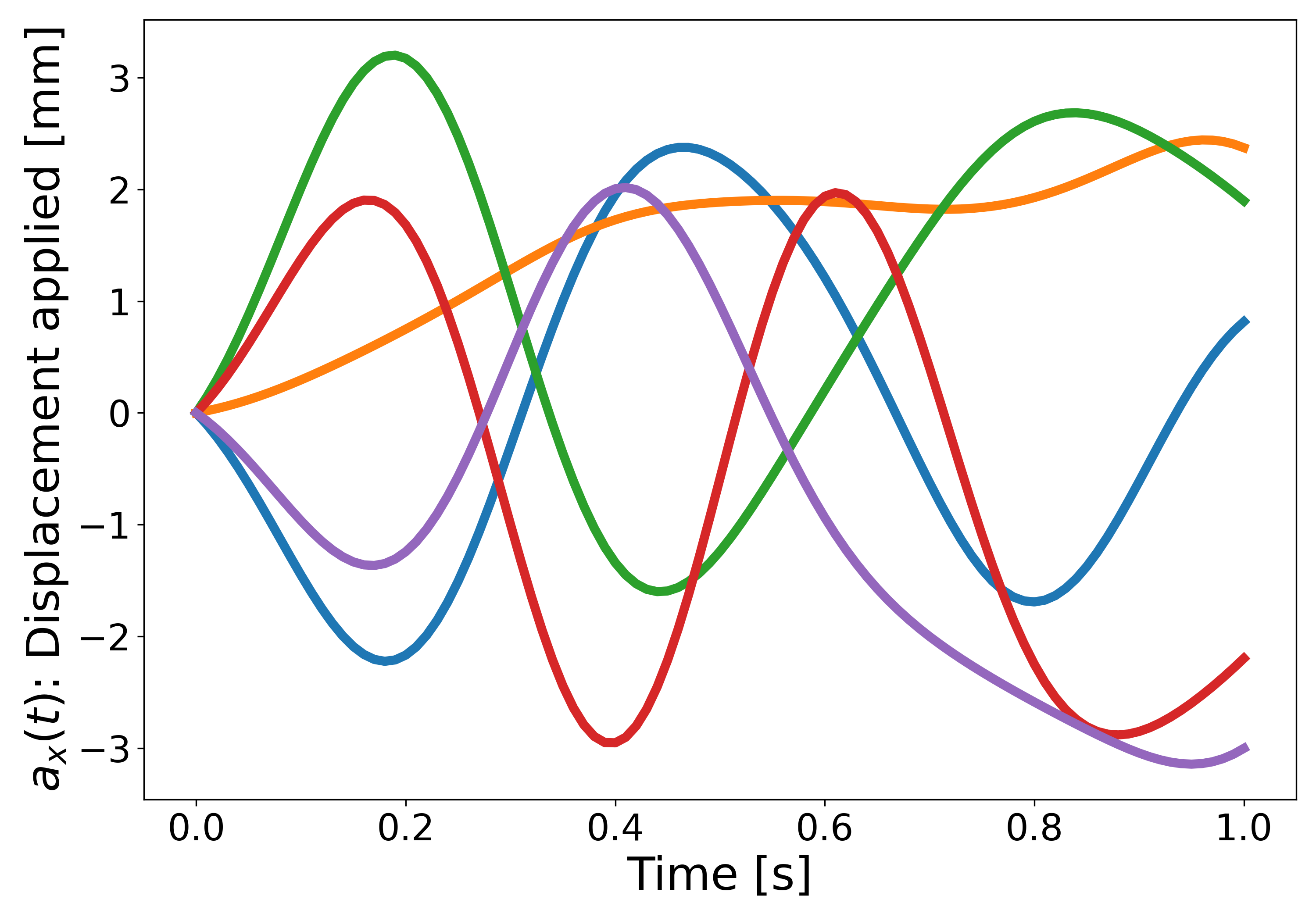}
        \subcaption{Five examples of boundary displacement conditions.}
        \label{fig:dogbone_load}
    \end{minipage}
    \caption{Specimen geometry and representative applied boundary conditions.}
    \label{fig:dogbone_sample_load}
\end{figure}

The second benchmark consists in-plane plastic deformation of a dogbone specimen with sequential loading applied through the simulation with FEA \cite{Abaqus2024}. The domain with the mesh grid that finite element (FE) simulations are represented in \fref{fig:dogbone_sample}. With the shape of a dogbone, it has overall length of $110$ mm, the grip width $30$ mm and reduced gauge width $20$ mm. The domain is discretized with 4756 linear, plane-stress elements (four-node quadrilaterals and three-node triangles) of unit thickness of $1$mm indicating plane-stress, yielding the mesh shown in the same figure. 

Quasi-static equilibrium in the absence of body forces is enforced through the equations in \eref{dogbone_governing_eq}, while $\boldsymbol{n}, \boldsymbol{\tilde{t}}$ and $\boldsymbol{\tilde{u}}$ representing outward unit normal vector at boundary, displacement constraints, and traction loads, respectively.

\begin{equation}
\begin{aligned}
    \nabla\!\cdot\!\boldsymbol{\sigma} &= \mathbf{0} \quad \text{in} \quad  \Omega, \\
    \boldsymbol{\sigma} \cdot \boldsymbol{n} &= \boldsymbol{\tilde{t}} \quad \text{in} \quad  \Omega_t \\
    \boldsymbol{u} &= \boldsymbol{\tilde{u}} \quad \text{in} \quad  \Omega_u
\end{aligned}
\label{dogbone_governing_eq}
\end{equation}

Small-strain kinematics are assumed, so that the total strain tensor is the symmetric gradient of the displacement field and is additively decomposed into elastic and plastic parts \cite{simo1998computational}:

\begin{equation}
    \boldsymbol{\epsilon}=\boldsymbol{\epsilon}^{\mathrm elastic}+\boldsymbol{\epsilon}^{\mathrm plastic}
\end{equation}

Also, since small deformation was assumed, the total strain can be derived  as:
\begin{equation}
    \boldsymbol{\epsilon}=\frac{1}{2}\left(\nabla \boldsymbol{u}+\nabla \boldsymbol{u}^T\right).
\end{equation}

Furthermore, letting the elastic modulus to $E$ and the Poisson ratio to $\nu$, plane stress condition gives the following constitutive equation:
\begin{equation}
    \left[\begin{array}{l}
\sigma_{11} \\
\sigma_{22} \\
\sigma_{12}
\end{array}\right]=\left[\begin{array}{ccc}
\frac{E}{1-\nu^2} & \frac{\nu E}{1-\nu^2} & 0 \\
\frac{\nu E}{1-\nu^2} & \frac{E}{1-\nu^2} & 0 \\
0 & 0 & \frac{E}{2(1+\nu)}
\end{array}\right]\left[\begin{array}{l}
\epsilon_{11} \\
\epsilon_{22} \\
\epsilon_{12}
\end{array}\right]
\end{equation}

Elastic response follows Hooke’s law for an isotropic, plane-stress solid, whereas plastic flow obeys $J_2$ theory \cite{lubliner2008plasticity} with linear isotropic hardening:

\begin{equation}
    \sigma_y\left(\bar{\epsilon}_p\right)=\sigma_{y 0}+H \bar{\epsilon}_p ,
\end{equation}

where $\sigma_{y0}$ is the initial yield stress and $H$ is the hardening modulus adopted. Detailed material properties used in this work are listed in \tref{dogbone_mat_props}.

\begin{table}[h!]
\centering
\caption{Elastic-plastic material parameters for the computational model}
\label{dogbone_mat_props}
\begin{tabular}{|c|c|c|c|c|}
\hline
\textbf{Property} & $E$ [MPa] & $\nu$ [-] & $\sigma_{y0}$ [MPa] & $H$ [MPa] \\
\hline
\textbf{Value} & $2.09 \times 10^5$ & 0.3 & 235 & 800 \\
\hline
\end{tabular}
\end{table}

All nodes on the left boundary are fixed, and a time-dependent horizontal displacement $a_x(t)$ was prescribed on the right boundary. The loading path is defined by six control points: the endpoints at $t=0$ and $t=1$ s plus four interior points whose times are drawn uniformly from $(0.1,0.9)$ s. The corresponding displacements are sampled so that the nominal axial strain never exceeds $5\%$. A radial-basis interpolant yields a smooth history between the control points; an example is shown in \fref{fig:dogbone_load}. A total of 9314 unique loading histories generated in this manner are solved with Abaqus/Standard. For each case we record the equivalent (von Mises) stress video:
\begin{equation}
    \bar{\sigma}=\sqrt{\sigma_{11}^2+\sigma_{22}^2-\sigma_{11}\sigma_{22}+3\sigma_{12}^2},
\end{equation}
which constitutes the ground-truth target for training and evaluation of the surrogate models. To put in later models, the domain of the dogbone specimen was interpolated into 128*128 square domain.

\subsection{Neural network framework}
\label{sec:architecture_design}

\subsubsection{S-DeepONet for Baseline Field Prediction}
\label{sec:sdon_architecture_design}

Since the dataset created in \ref{sec:ldc_training_data_generation}, \ref{sec:dogbone_training_data_generation} highly depend on time-history inputs, we adopt neural operator S-DeepONet \cite{he2024sequential}. 

\begin{figure}[h!]
    \centering
         \includegraphics[width=1.0\textwidth]{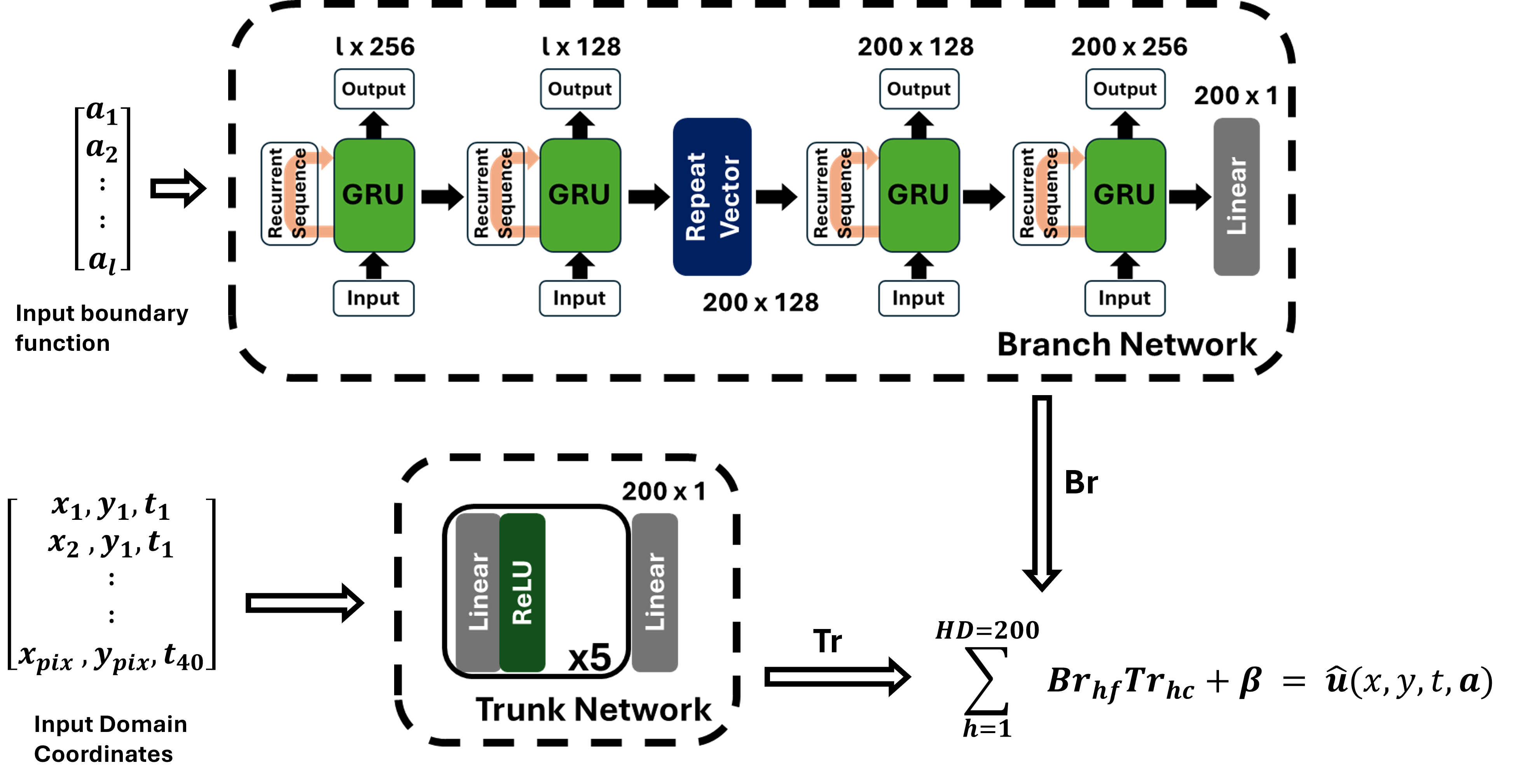}
    \caption{S-DeepONet architecture tailored for video data. $l$ denotes the input boundary function length, pix represents the pixel in each direction. (64 for lid driven cavity problem, 128 for plastic deformation of dogbone specimen.) HD, the hidden dimension was set to 200 in this work.}
    \label{sdon_model}
\end{figure}

Comprised of GRU models \cite{cho2014learning}, the model is able to capture sequential information in time series signal input. Our implementation of S-DeepONet inherits the work of \cite{he2024sequential}, preserving using two separate neural networks: branch and trunk. Detailed network architecture is shown in \fref{sdon_model}.
The branch network receives sequential data such as the lid velocity from \ref{sec:ldc_training_data_generation}, and discretized load history from \ref{sec:dogbone_training_data_generation}. The time-dependent input gets processed with 4 GRU units inside the network and becomes a latent vector of length hidden dimension (HD) which will later inner-producted with the trunk output. On the other hand, the trunk network takes the coordinates of the domain of the problem. Since the data is time-evolving 2D images (or simply videos), we put (x, y, t) coordinate to the trunk network. Comprised of 6 FNN layers with ReLU in between, a coordinate given to the trunk gets transformed into a latent vector of length HD also. Then, the two latent vectors, each from the branch network and the trunk network gets batched-inner-product, implemented via an Einstein-summation call, to represent the solution value for a given coordinate. The formulation is presented at \eref{sdon_equation}.

\begin{equation}
\hat{u}(x,y,t,\boldsymbol{a}) = \sum_{i=1}^{HD} Br_i(\boldsymbol{a}) Tr_i(x, y, t) + \beta
\label{sdon_equation}
\end{equation}

In \eref{sdon_equation}, $\hat{u}$ represents the predicted solution field. It takes four inputs $x, y, t, \boldsymbol{a}$. The first three parameters $x, y, t$ are spatio-temporal coordinate of the solution field which are put also into the trunk network denoted as $Tr$, and the fourth parameter $\boldsymbol{a}$ is the sequential function vector input that also goes into the brach network $Br$. Both branch and the trunk network will output a vector of length HD. Finally $\beta$ is a bias which also is a trainable parameter defined inside the network. Throughout the remainder of the paper we treat the video $\hat{u} (\cdot, \bm{a})$ as a coarse, physics-consistent prior and denote it by $\mathbf{x^{prior}}$. In the next section we show how this prior is incorporated to condition a video diffusion corrector that restores the high-frequency content absent from S-DeepONet.

\subsubsection{Video Diffusion for Prior Refinement}
\label{sec:imagen_video_architecture_design}

\paragraph{Elucidated Video Diffusion Model.}

To recover fine-scale details from S-DeepONet prior, we adopt Imagen video \cite{ho2022imagenvideo} implemented here with the Elucidated Diffusion Model (EDM) sampler of \cite{karras2022elucidating}. The model is a classifier-free-guided, conditional diffusion model. The network eventually generates the refined solution video (e.g. stream function field or von Mises stress field) given two conditions: the S-DeepONet prior and the same input function that feeds the branch network of S-DeepONet.

\begin{figure}[h!]
    \centering
         \includegraphics[width=1.0\textwidth]{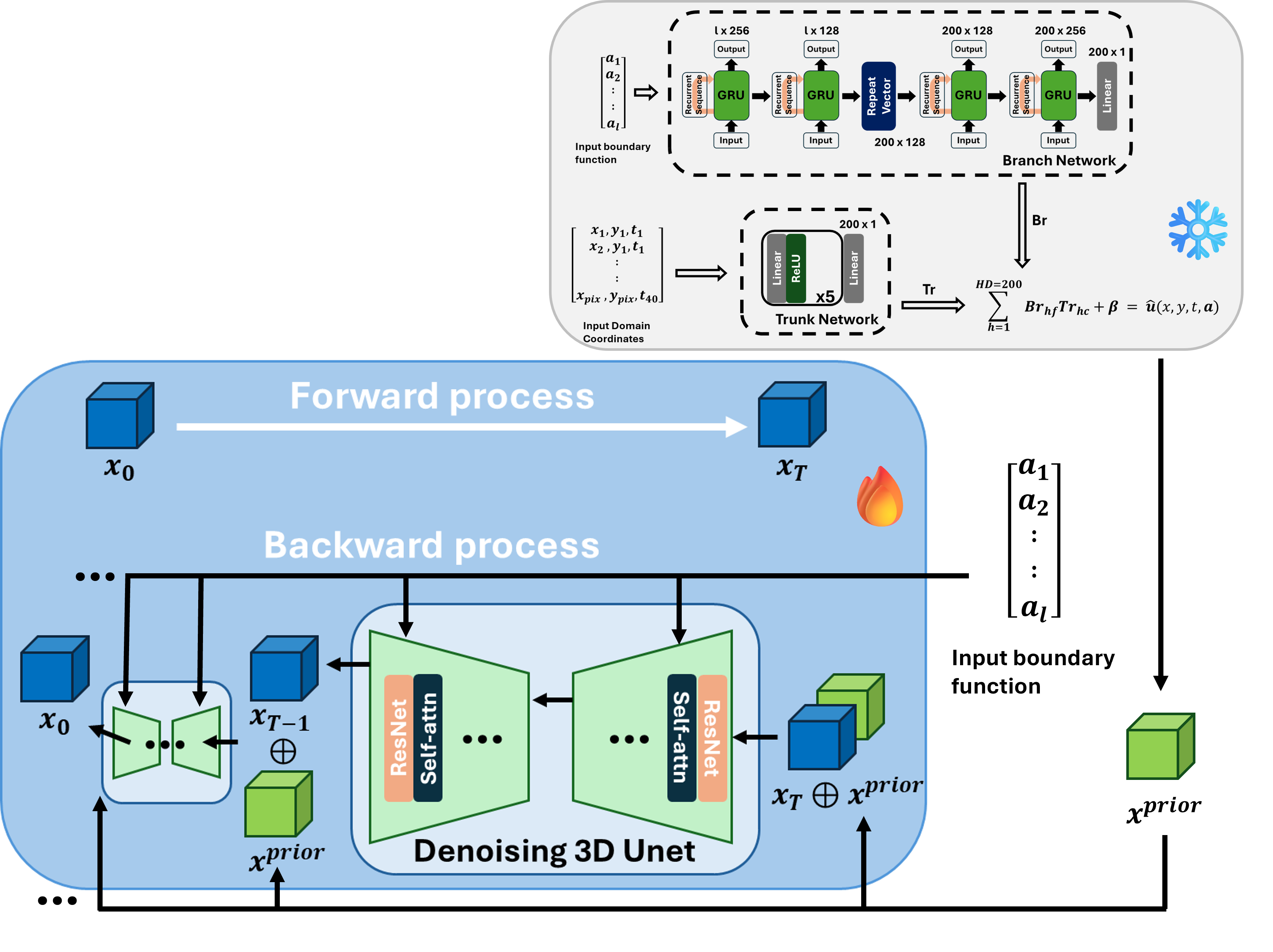}
    \caption{Video diffusion model architecture. We freeze the trained S-DeepONet model and obtain prior video ($x^{prior}$). Re-using the input boundary function as a condition along with the prior video, only the video diffusion model was trained. $\oplus$ denotes concatenation in the channel axis.}
    \label{imagen_video_model}
\end{figure}

Following the stochastic differential equation formulation of \cite{song2021score}, a clean video $\boldsymbol{\mathbf{x_0}}$ is progressively corrupted by Gaussian noise:
\begin{equation}
    \boldsymbol{\mathbf{x}}(t)=\boldsymbol{\mathbf{x_0}}+\sigma(t) \boldsymbol{\epsilon}, \qquad \boldsymbol{\epsilon} \sim \mathcal{N}\left(\mathbf{0},  \mathbf{I}\right),
\label{forward_sde}
\end{equation}

where $\boldsymbol{\mathbf{x}}(t)$ denotes the noised sample at noise timestep $t$, $\boldsymbol{\mathbf{x_0}}$ is the original clean image, $\sigma(t)$ is the predefined noise schedule, and $\boldsymbol{\epsilon}$ is the sampled noise from the Gaussian distribution to be added. Then, EDM solves the reverse (denoising) dynamics with the following stochastic differential equation in \eref{backward_sde}:

\begin{equation}
    \mathrm{d} \boldsymbol{x}=-\dot{\sigma}(t) \sigma(t) \nabla_{\boldsymbol{x}} \log p(\boldsymbol{x} ; \sigma(t)) \mathrm{d} t - \beta(t) \sigma(t)^2 \nabla_{\boldsymbol{x}} \log p(\boldsymbol{x} ; \sigma(t)) \mathrm{d} t+\sqrt{2 \beta(t)} \sigma(t) \mathrm{d} \omega_t
\label{backward_sde}
\end{equation}

with $\beta(t)=\dot\sigma(t)/\sigma(t)$ during churn steps ($\gamma\!>\!0$ in Algorithm 2 of \cite{karras2022elucidating}) and $\beta(t)=0$ during deterministic Heun steps. The score $\nabla_x\log p(x;\sigma)$ is estimated by the diffusion network and pre/post conditioned with the scaling factors $c_{\mathrm{in}},c_{\mathrm{skip}},c_{\mathrm{out}}$ described in Eq.~(7) of the same paper.

\noindent
\textbf{Note.} Our notation follows the conventional choice $x_0\!=\!$ clean data. In Algorithm 2 of \cite{karras2022elucidating}, the indices are reversed ($x_0$ = max-noise, $x_N$ = clean). The underlying mathematics, however, is identical.

\paragraph{Conditioning.}

As a denoiser, we employ a depth-4 encoder–decoder 3D U-Net equipped with spatio–temporal convolutions and attention blocks. Unlike latent diffusion, the network operates directly in pixel space to avoid compression artifacts that could obscure physical details \cite{wallace2023edict}. Formally, the model learns the conditional distribution
\begin{equation}
  p_\theta\!\bigl(\mathbf{x_0} \,\big|\, 
        \bm a,\;
        \mathbf x^{\text{prior}}),
\end{equation}
where $\mathbf{x_0}$ is the clean target video, $\bm{a}$ the input lid-velocity/load function, and \(\mathbf x^{\text{prior}}\) the S-DeepONet prior. To guide the model with these conditions, we inject two signals ($\bm{a}$ and $\mathbf x^{\text{prior}}$) through separate pathways as detailed below.

\begin{enumerate}[label=(\roman*)]
    \item Physics-input FiLM modulation.  \\
          The lid-velocity / load function $\bm{a}$, identical to the branch input of S-DeepONet, is passed through two linear layers with SiLU activation, yielding a latent code $\gamma(\bm{a})$.  
          This latent code modulate every feature map channel-wise via FiLM \cite{perez2018film}, continuously steering the U-Net’s internal representation toward the current boundary/loading regime.
    \item Prior-conditioned concatenation. \\
          The coarse S-DeepONet prediction $\mathbf{x}^{\text{prior}}$ is appended to the noisy input at every diffusion step:
          \begin{equation}
              \mathbf{x}_t\leftarrow\mathbf{x}_t \oplus \mathbf{x}^{\text{prior}},
          \end{equation}
          supplying a low-frequency anchor, guiding the denoiser toward a more favorable basin of the loss landscape. (Here, $\oplus$ denotes channel-wise concatenation.) This “warm-start’’ allows the U-Net to focus on correcting high-frequency discrepancies rather than reinventing the entire solution field. For the plastic deformation of the dogbone problem, we introduce an additional binary mask $ \mathbf{m}^{\text{dogbone}}$ that distinguishes material from void. This geometric constraint is concatenated in the same manner, yielding
          \begin{equation}
              \mathbf{x}_t\leftarrow\mathbf{x}_t \oplus \mathbf{x}^{\text{prior}} \oplus \mathbf{m}^{\text{dogbone}}.
          \end{equation}
        Incorporating the mask prevents the denoiser from hallucinating stress or strain outside the physical domain.
\end{enumerate}

\paragraph{Time-wise focal loss.}

Although the 3D U-Net is trained with a pixel-wise reconstruction objective, sharp shear layers or stress singularities which transform rapidly may occupy only a few frames yet dominate the physical fidelity. To prevent these “difficult’’ frames from being drowned out by the many easy ones, we introduce a time-wise focal loss inspired by \cite{Lin_2017_ICCV}, but applied along the temporal axis only.

Having the shape of the video (B, C, T, H, W), let the pixel-wise training loss returned by the diffusion model be
\begin{align}
\mathcal{L}_{b,t}= \frac{1}{C \cdot H \cdot  W}\sum_{c,h,w}
\lvert x_{b,c,t,h,w}-\hat x_{b,c,t,h,w}\rvert^2,
\end{align}
averaged over channels and spatial dimensions. We first obtain the mean loss for each frame,
\begin{align}
\ell_t = \frac{1}{B}\sum_{b}\mathcal{L}_{b,t},
\end{align}
and its global mean
\begin{align}
\bar{\ell}=\frac{1}{T}\sum_{t=1}^{T}\ell_t.
\end{align}
The focal weight for frame $t$ is then defined as
\begin{equation}
    w_t=\Bigl(\frac{\ell_t}{\bar{\ell}+{\varepsilon}}\Bigr)^{\xi}, \qquad \xi=2,
    \label{eq:focal_weight}
\end{equation}
where $\varepsilon=10^{-8}$ prevents division by zero.
The final training objective is
\begin{equation}
    \mathcal{L}_{\text{focal}}=
    \frac{\sum_{t=1}^{T} w_t\,\ell_t}{\sum_{t=1}^{T} w_t+{\varepsilon}}.
    \label{eq:focal_loss}
\end{equation}

This re-weighting strategy keeps the loss on the same overall scale via normalization yet automatically amplifies the frames whose per-frame error $\ell_t$ exceeds the running average $\bar{\ell}$ by the scale of $\xi$. As a result, the network allocates more gradient budget to frames that exhibit relatively larger errors that are critical for downstream physical analysis but rare in time. $\xi$ is a tunable hyperparameter, $\xi=2$ was taken empirically in this work. Detailed network architecture is shown in \fref{imagen_video_model}.

\subsubsection{Target formulations for Video Diffusion: Full solution field vs. Residual}
We experiment two cases each having different targets for the video diffusion stage introduced in \ref{sec:imagen_video_architecture_design}: the full solution field, and the residual field. 

\begin{itemize}
\item \textbf{Full solution field target} – the network learns the entire true solution video
      $\mathbf{x^{GT}}$.
\item \textbf{Residual field target} – the network learns only the discrepancy $\mathcal{R} = \mathbf{x^{GT}} - \mathbf{x^{prior}}$ between the ground truth and the
      S‑DeepONet prior $\mathbf{x^{prior}}$.
\end{itemize}

Predicting the entire solution video $\mathbf{x^{GT}}$ (Ground Truth), although conceptually straightforward, forces the diffusion model to reproduce both coarse rough structures that the operator prior already approximates well and high‐fidelity details that it misses. Based on multiple reports on using residual information \cite{yue2023resshift, mardani2025residual, liu2024residual, chen2023resgrad, choukroun2022denoising}, we hypothesize that training on the residual $\mathcal{R}$ yields a lower‐entropy target, accelerates convergence, and improves accuracy because the network can focus its capacity on the under‐resolved fine scales while the coarse field is “given for free’’ by S‐DeepONet. During training all residual frames are linearly mapped to the interval $[-1, 1]$ to match the dynamic range expected by the diffusion backbone.
\begin{comment}
\begin{equation}
    \widetilde{\mathcal{R}}
       = \frac{( \mathcal{R}-\mathcal{R}_{\min} )}{(\mathcal{R}_{\max}-\mathcal{R}_{\min})}\times2-1 .
\end{equation}
\end{comment}

All other components such as conditioning channels, U‑Net architecture, loss (\ref{sec:imagen_video_architecture_design}, \eref{eq:focal_loss}), and training schedule remain identical between the two experiments. At inference,
the residual variant reconstructs the solution by:
\begin{equation}
\begin{aligned}
    \mathbf{\hat{x}} &= \hat{\mathcal{R}}_{unnormed} + \mathbf{x^{prior}}, \quad \text{where} \\
    \hat{\mathcal{R}}_{unnormed} &= \frac{(\hat{\mathcal{R}}+1)}{2} \times(\mathcal{R}_{max} - \mathcal{R}_{min}) + \mathcal{R}_{min}.
\end{aligned}
\label{residual_reconstruction}
\end{equation}

In \eref{residual_reconstruction}, $\mathbf{\hat{x}_0}$ is the predicted entire solution field, $\hat{\mathcal{R}}_{unnormed}$ is the residual field but unnormed to solution field scale. $\hat{\mathcal{R}}$ is the raw predicted residual field directly out from the diffusion model, $\mathcal{R}_{max}, \mathcal{R}_{min}$ are the global maximum and minimum residual values measured on the training set. $\hat{\mathcal{R}}_{unnormed}$ was calculated from $\hat{\mathcal{R}}$ to scale back to the solution field so that it can be added to the prior to get a refined field. Section \ref{sec:results} will discuss results of these two experiments more in detail - full solution field as a target, and a residual field as a target.

\section{Results and discussion}
\label{sec:results}

Datasets from physics-based simulation were generated with Abaqus/Standard \cite{Abaqus2024} and ANSYS Fluent \cite{AnsysFluent2024} running on AMD EPYC 7763 (“Milan”) CPU cores. All deep-learning experiments including the S-DeepONet and video diffusion model were implemented in PyTorch and trained on NVIDIA GH200 GPU nodes of the DeltaAI cluster.

To evaluate model performance, two metrics were used to measure accuracy between the true solution field and the predicted solution field: the relative mean $L_2$ error (Rel. $L_2$ error) and relative mean absolute error (RMAE). They are applied on the test set obtained by the 80-20 train-test split.
\begin{equation}
\begin{aligned}
\text{Relative } L_2 \text{ error} &= \frac{1}{T} \sum_{t=1}^{T}\frac{|s_{True}^{(t)} - s_{Pred}^{(t)}|_2}{|s_{True}^{(t)}|_2}  \\
\text{Relative mean absolute error} &= \frac{1}{T} \sum_{t=1}^{T}\frac{|s_{True}^{(t)} - s_{Pred}^{(t)}|_1}{|s_{True}^{(t)}|_1}
\label{eval_metrics}
\end{aligned}
\end{equation}

\eref{eval_metrics} defines an error between the ground truth and the predicted solution field for one simulation, where $T, s_{True},$ and $s_{Pred}$ denote the number of timesteps, ground truth solution field obtained by CFD or FEA, and predicted solution field respectively.

To see prior utilization effect in \ref{sec:imagen_video_architecture_design}, we conduct an \emph{ablation study} built around four models:

\begin{enumerate}[label=(\roman*)]
    \item \textbf{S-DeepONet} — the operator network acting \emph{alone}, providing a pure–prior reference.  
    \item \textbf{VD-NP} — the standalone video diffusion model with \emph{no} prior, showcasing only generative capacity with no operator learning incorporated.
    \item \textbf{VD-PC-D} — video diffusion with prior conditioning that \emph{directly} output the full solution field, measuring how much the prior improves the result.  
    \item \textbf{VD-PC-R} — video diffusion with prior conditioning that predicts only the \emph{Residual} field, allowing the model to focus its capacity on the fine-scale corrections the prior cannot capture.  
\end{enumerate}

By comparing (i) and (ii) we gauge the standalone strengths of operator learning versus diffusion. Contrasting (ii) with (iii) reveals the incremental benefit of feeding the prior into diffusion, and comparing (iii) with (iv) tests whether residual learning more efficiently allocates model capacity than direct prediction.  
These four variants therefore span all meaningful combinations of “prior/no-prior’’ and “direct/residual’’ strategies, forming a minimal yet comprehensive grid for disentangling the impact of each design decision. For brevity, the abbreviated labels above reappear consistently in subsequent figures, tables, and discussions.

%To see prior utilization effect in \ref{sec:imagen_video_architecture_design}, we conduct an \emph{ablation study} built around four models: the standalone S-DeepONet which is the prior, the standalone video diffusion model with no prior used (VD-NP), video diffusion model with prior conditioning and predicting full solution field directly (VD-PC-D), and video diffusion model with prior conditioning generating residual field (VD-PC-R). These names appear repeatedly hereafter in figures, tables, and in texts. They systematically serve ablation study.

\subsection{Predicting lid driven cavity flow}

\paragraph{Quantitative result analysis}

\begin{table}[h!]
\centering
\caption{Overall statistics of error metrics for the stream function field in the lid driven cavity benchmark.}
\begin{tabularx}{\textwidth}{ 
  >{\centering\arraybackslash}X 
  >{\centering\arraybackslash}X 
}
\begin{tabular}{c|c|c|c|c}
\multicolumn{5}{c}{\textbf{Lid driven cavity flow}} \\ \hline
\textbf{}  & \textbf{S-DeepONet} & \textbf{VD-NP} & \textbf{VD-PC-D} & \textbf{VD-PC-R} \\ \hline
Mean Rel. $L_2$      & 4.58\%       & 8.24\%      & 1.11\%      & 0.829\%    \\ 
Mean RMAE            & 4.33\%       & 8.25\%      & 1.16\%      & 0.808\%      
\end{tabular}
\end{tabularx}
\label{tab:ldc_error_overall}
\end{table}

\tref{tab:ldc_error_overall} presents the mean errors defined in \eref{eval_metrics} to show overall model performance. Both the mean Rel. $L_2$ error and the mean RMAE shows that utilizing the S-DeepONet information to the video diffusion model setting residual field as a target (VD-PC-R) provides the best result, meaning it generates the most accurate solution field video across different models.

\begin{figure}[h!] 
    \centering
    \begin{minipage}{0.48\textwidth}
        \centering
        \includegraphics[width=\textwidth]{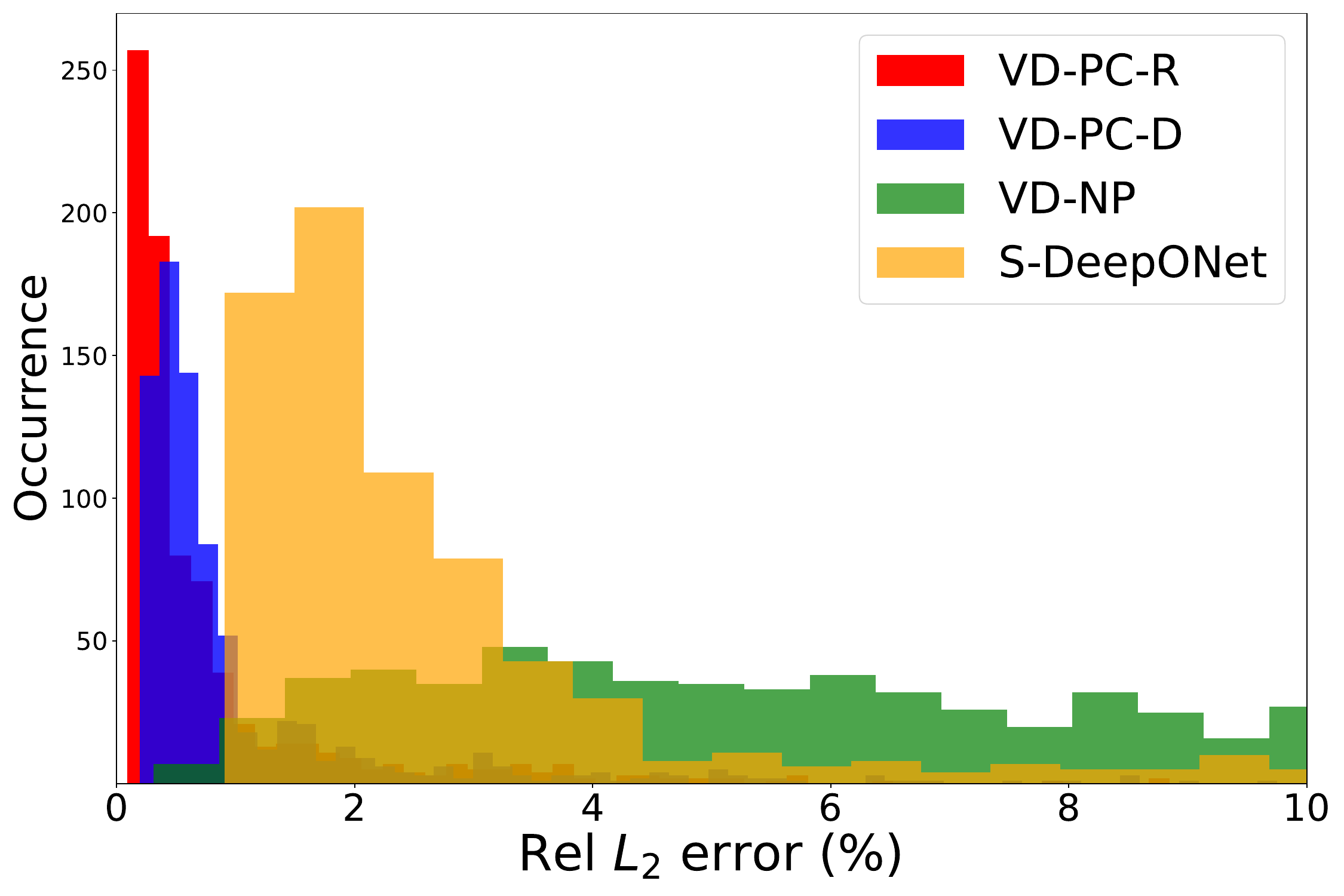}
        \caption{Histogram of Rel. $L_2$ error for lid driven cavity flow example.}
        \label{fig:ldc_l2_hist}
    \end{minipage}
    \hfill
    \begin{minipage}{0.48\textwidth}
        \centering
        \includegraphics[width=\textwidth]{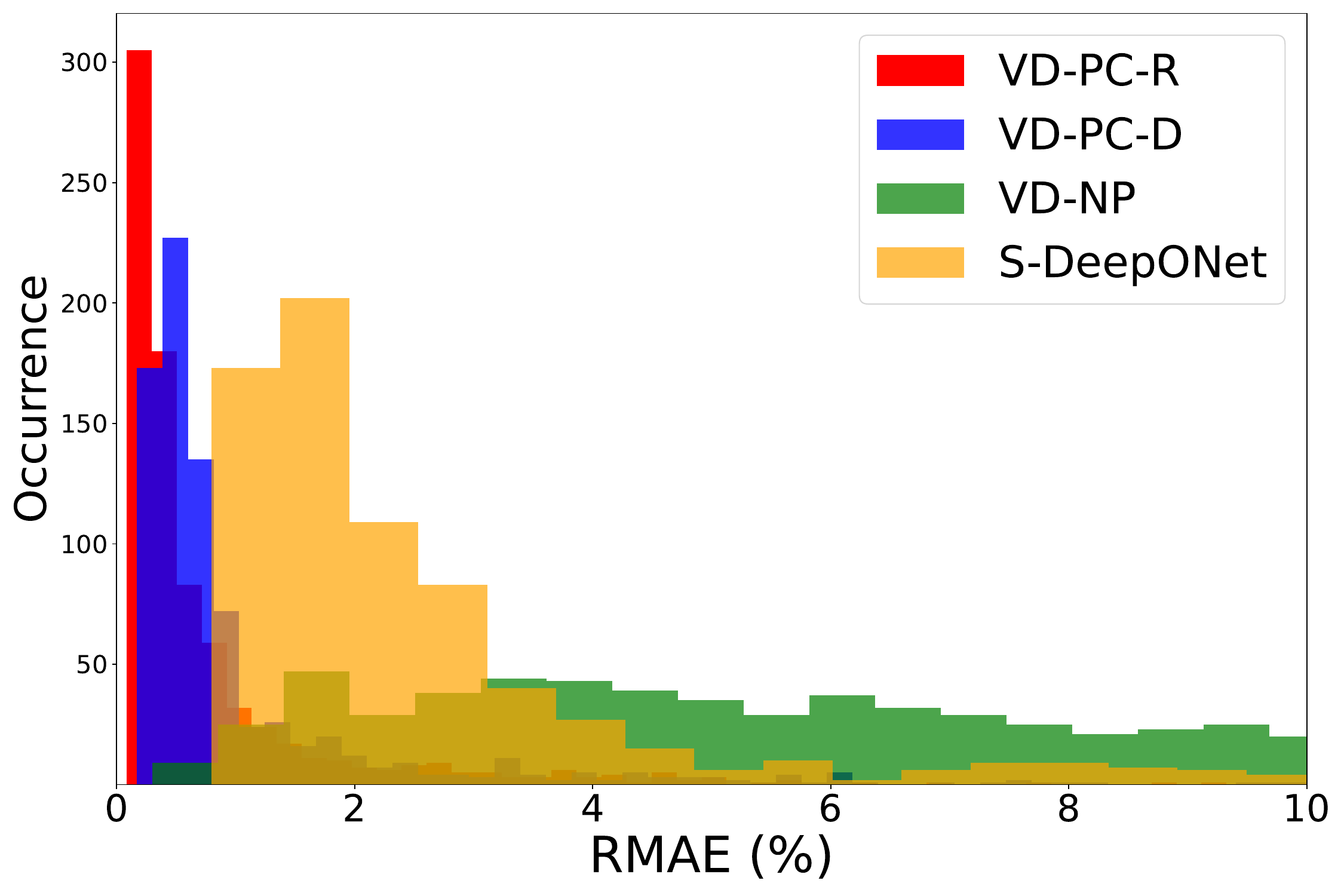}
        \caption{Histogram of RMAE for lid driven cavity flow example.}
        \label{fig:ldc_rmae_hist}
    \end{minipage}
\end{figure}

Key statistics of both the relative $L_2$ error and the relative mean absolute error over test cases are summarized in \tref{tab:l2_cavity_percentile} and \tref{tab:mae_cavity_percentile}. Across all percentiles, the S-DeepONet–conditioned diffusion models(VD-PC-D, VD-PC-R) delivers an order-of-magnitude improvement over the diffusion-only baseline and reduces the errors of the standalone S-DeepONet by roughly a factor of five. In particular, their median $L_2$ error falls below 1\%, compared with 4.09\% for S-DeepONet and 7.65\% for diffusion alone, while the worst case RMAE is cut from 13.2\% to 1.71\%. Histograms of the error distributions (\fref{fig:ldc_l2_hist} and \fref{fig:ldc_rmae_hist}) reveal that the hybrid model not only shifts the mass of the distribution toward lower errors but also compresses the long “heavy” tail exhibited by the two single-stage baselines.

\begin{table}[h!]
\centering
\caption{Percentile statistics of the relative $L_2$ error (\%) for the stream function field in the lid driven cavity benchmark.}
\begin{tabularx}{\textwidth}{ 
  >{\centering\arraybackslash}X 
  >{\centering\arraybackslash}X 
}
\begin{tabular}{c|c|c|c|c|c}
\multicolumn{6}{c}{\textbf{Rel. $L_2$ error by each percentile}} \\ \hline
\textbf{}  & \textbf{Best case} & $\mathbf{25^{th}}$ & $\mathbf{50^{th}}$ & $\mathbf{75^{th}}$ & \textbf{Worst case} \\ \hline
S-DeepONet                                    & 2.79\%       & 3.80\%      & 4.09\%      & 4.73\%     & 7.15\%    \\ 
VD-NP                    & 3.72\%       & 6.81\%      & 7.65\%      & 10.1\%     & 13.3\%    \\ 
VD-PC-D          & 0.623\%      & 0.810\%     & 0.975\%      & 1.23\%      & 1.93\%    \\
\textbf{VD-PC-R}  & 0.431\%      & 0.505\%    & 0.807\%      & 0.969\%     & 1.70\%
\end{tabular}
\end{tabularx}
\label{tab:l2_cavity_percentile}
\end{table}

\begin{table}[h!]
\centering
\caption{Percentile statistics of the relative mean absolute error (\%) for the stream function field in the lid driven cavity benchmark.}
\begin{tabularx}{\textwidth}{ 
  >{\centering\arraybackslash}X 
  >{\centering\arraybackslash}X 
}
\begin{tabular}{c|c|c|c|c|c}
\multicolumn{6}{c}{\textbf{RMAE by each percentile}} \\ \hline
\textbf{}  & \textbf{Best case} & $\mathbf{25^{th}}$ & $\mathbf{50^{th}}$ & $\mathbf{75^{th}}$ & \textbf{Worst case} \\ \hline
S-DeepONet                       & 2.56\%       & 3.63\%      & 3.93\%      & 4.42\%     & 6.79\%    \\ 
VD-NP                  & 3.74\%       & 6.79\%      & 7.67\%      & 10.2\%     & 13.2\%    \\ 
VD-PC-D  & 0.622\%      & 0.795\%     & 0.965\%     & 1.29\%     & 2.17\%    \\
\textbf{VD-PC-R}  & 0.405\%      & 0.485\%     & 0.775\%     & 0.921\%     & 1.71\%  
\end{tabular}
\end{tabularx}
\label{tab:mae_cavity_percentile}
\end{table}

\paragraph{Qualitative result analysis}

\begin{figure}[h!] 
    \centering
         \includegraphics[width=1.0\textwidth]{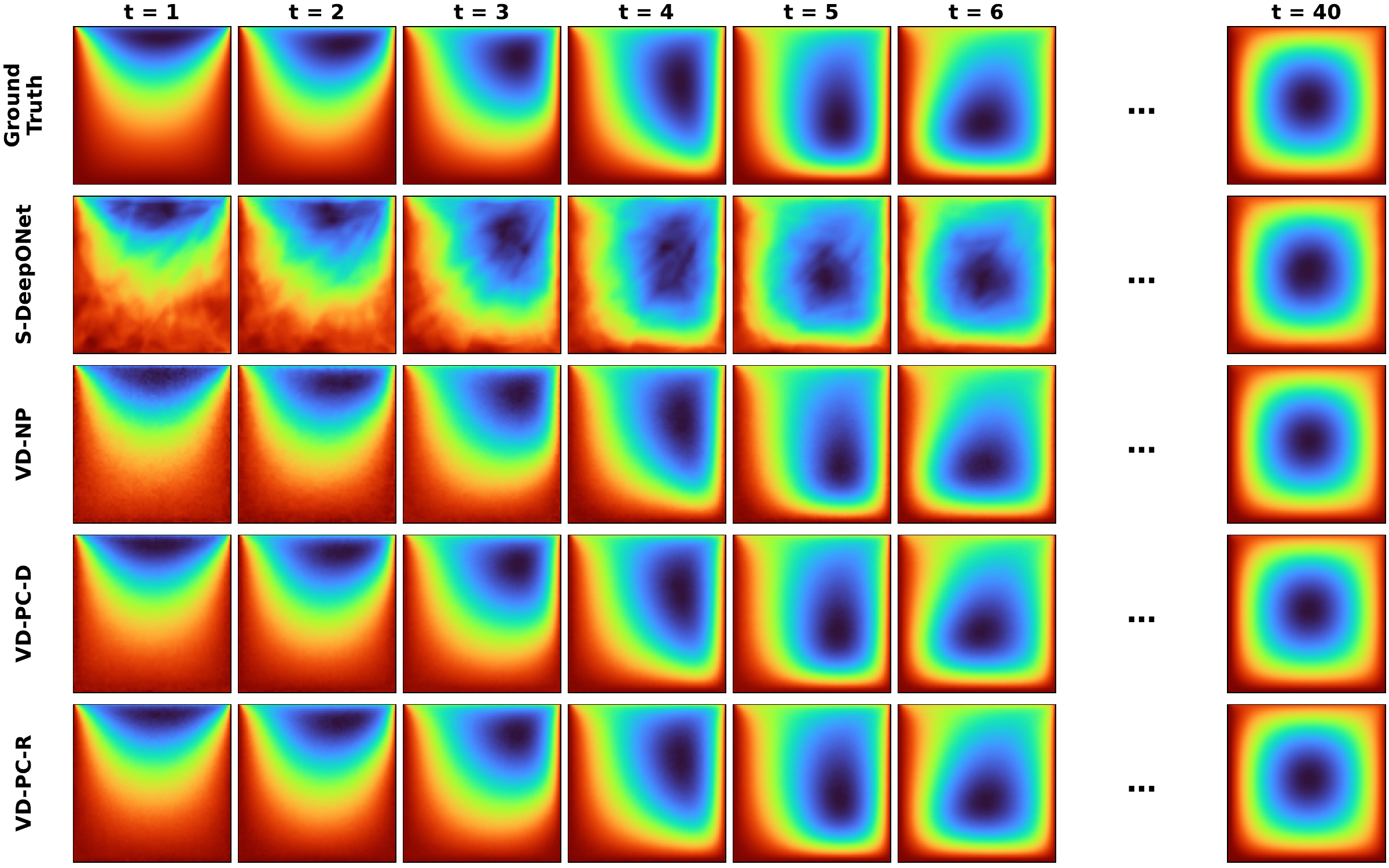}
    \caption{Stream function predictions for each model with true data for one representative sample. Temporal evolution of stream function field predictions across critical timesteps during simulation. Each row presents flow field predictions from different approaches: ground truth CFD videos, standalone S-DeepONet, standalone diffusion model, and the proposed two hybrid conditioned diffusion methods (both diffusion conditioned with prior - solution field and residual field variant). The selected timesteps (t = 1, 2, 3, 4, 5, 6) capture the progression from initial field evolution to final stage, highlighting how each model handles the most challenging transient phases where the temporal variations are most pronounced.}
    \label{ldc_preds}
\end{figure}

\begin{figure}[h!]
   \centering
   \begin{minipage}{0.49\textwidth}
       \centering
       \includegraphics[width=\textwidth]{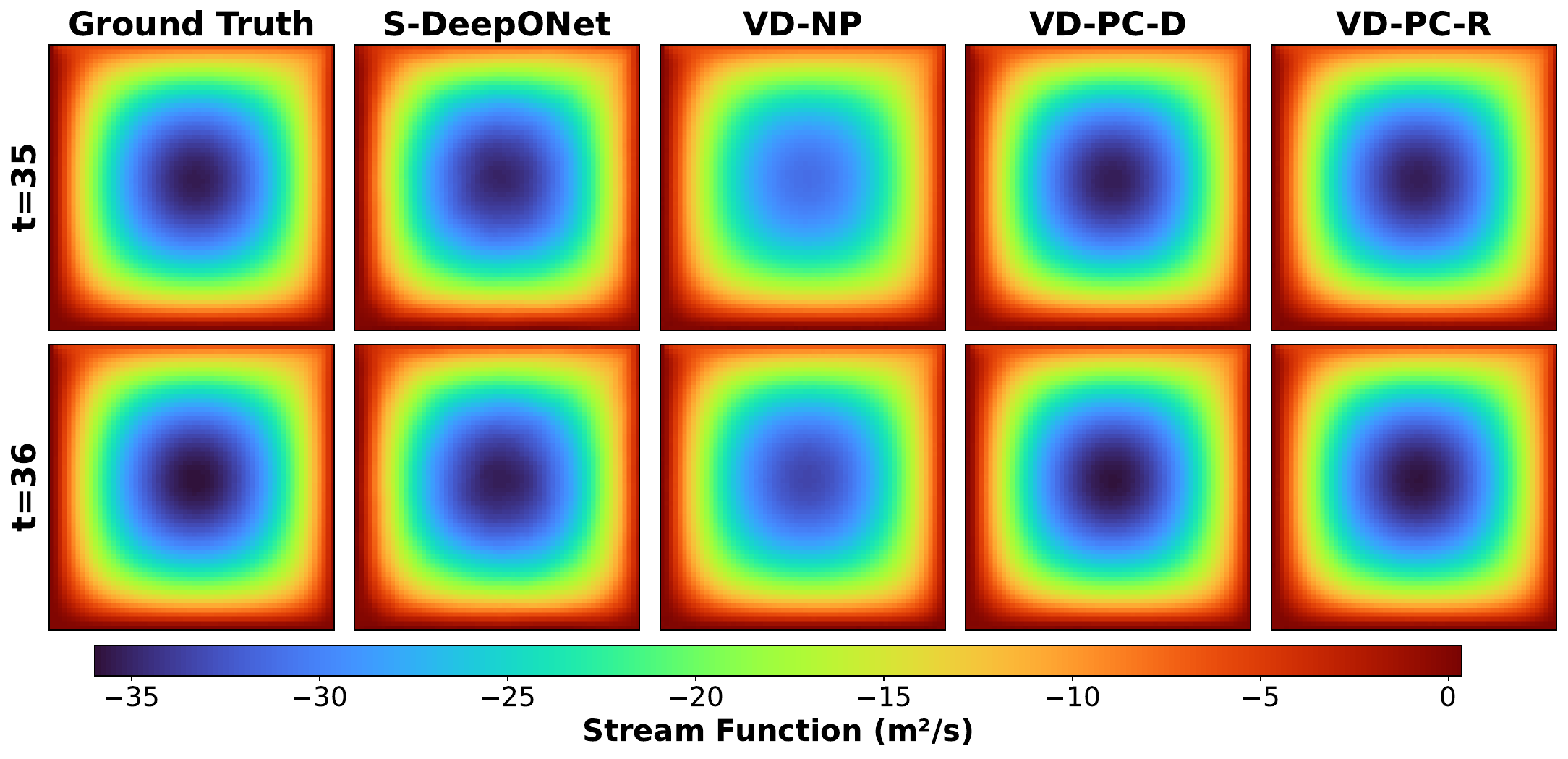}
       \subcaption{Stream function comparison}
       \label{ldc_quality_best_full_field}
   \end{minipage}
   \begin{minipage}{0.49\textwidth}
       \centering
       \includegraphics[width=\textwidth]{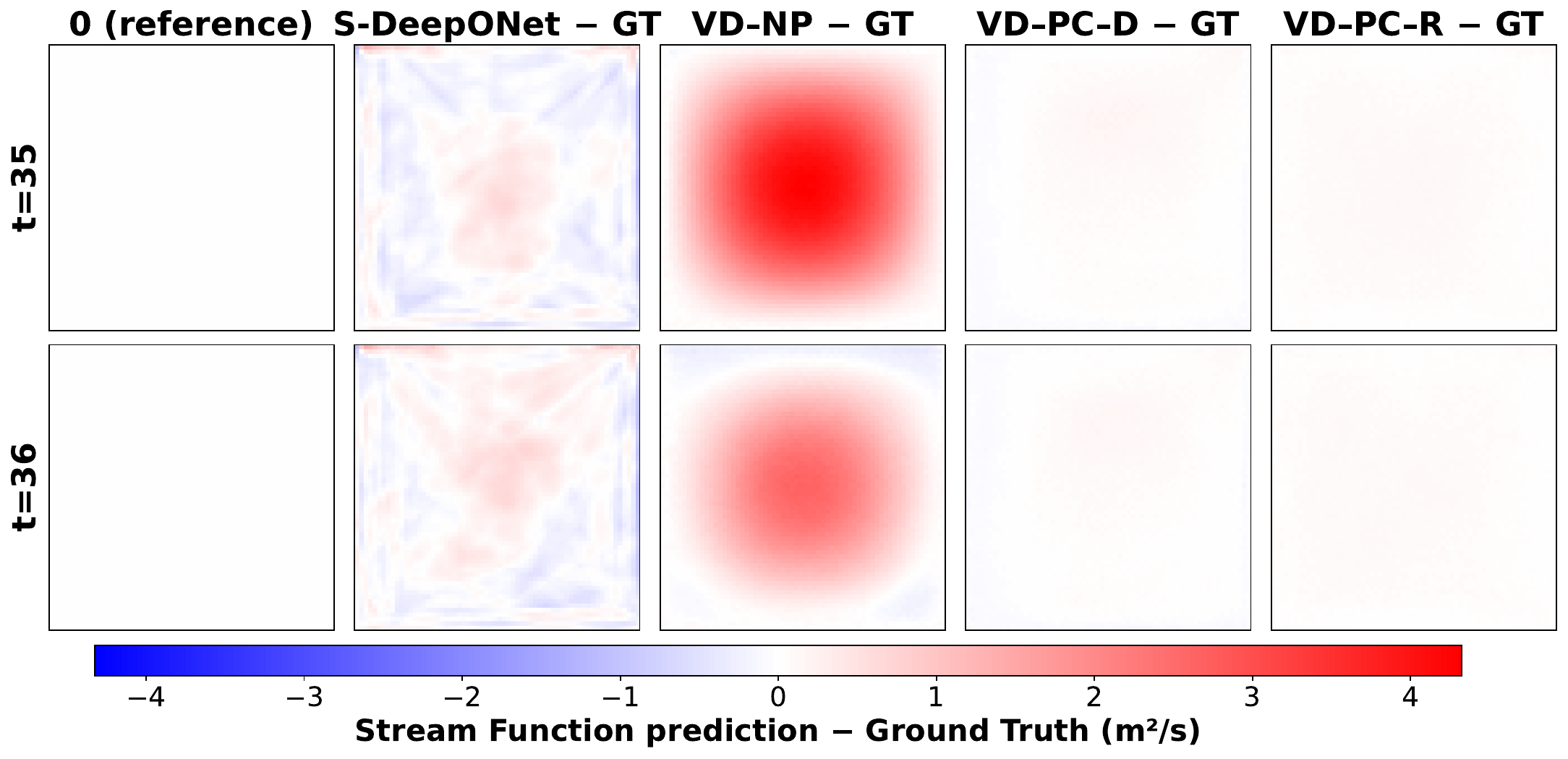}
       \subcaption{Residual comparison}
       \label{ldc_quality_best_residual}
   \end{minipage}
   \caption{Best-performing S-DeepONet case on Rel. $L_2$ error, qualitative comparison. Results are shown at a selected timestep that illustrates typical model performance. Residuals are computed as the difference between each model's prediction and the ground truth (GT) solution.}
   \label{ldc_quality_best}
\end{figure}

\begin{figure}[h!]
   \centering
   \begin{minipage}{0.49\textwidth}
       \centering
       \includegraphics[width=\textwidth]{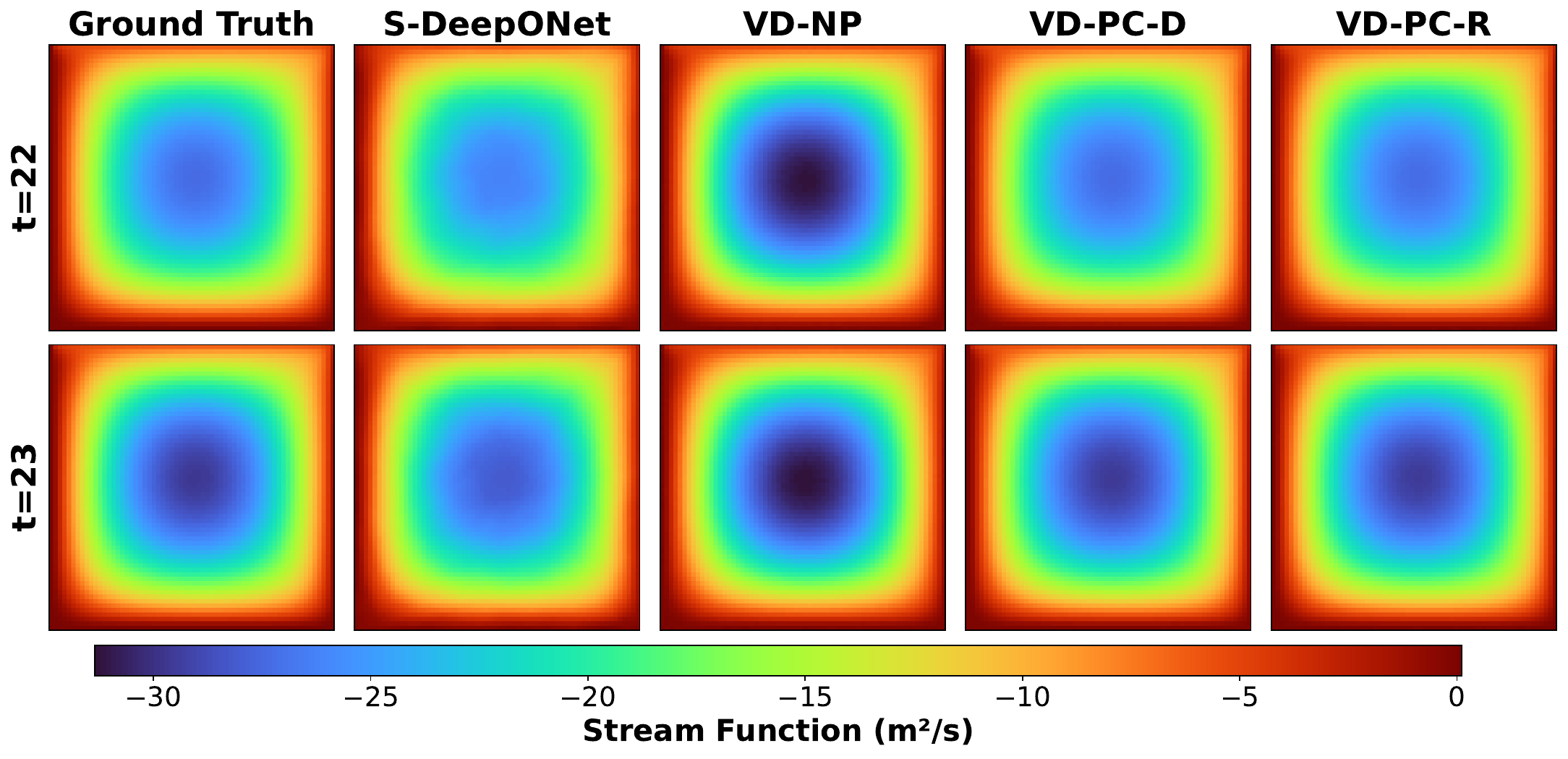}
       \subcaption{Stream function comparison}
       \label{ldc_quality_median_full_field}
   \end{minipage}
   \begin{minipage}{0.49\textwidth}
       \centering
       \includegraphics[width=\textwidth]{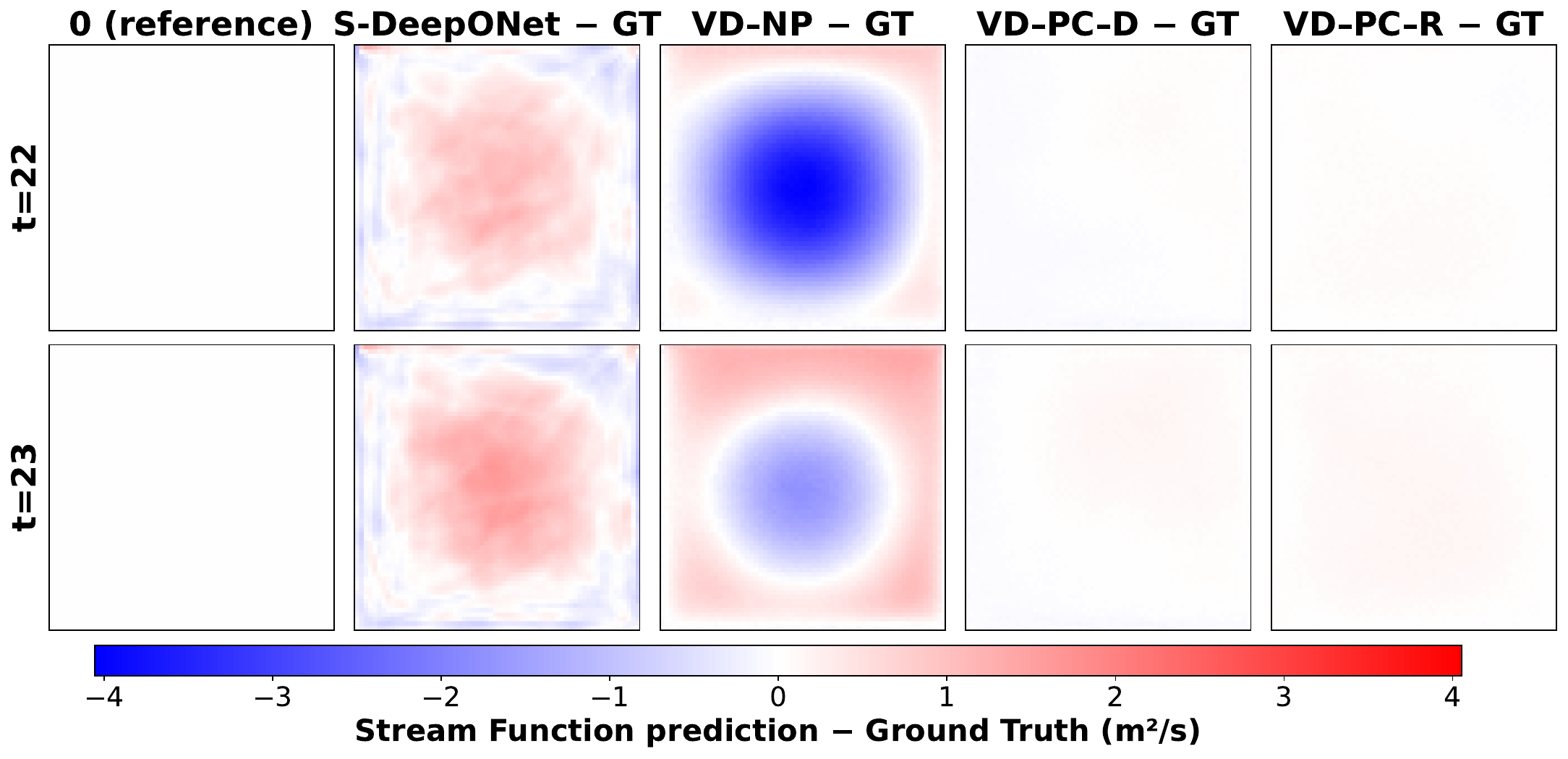}
       \subcaption{Residual comparison}
       \label{ldc_quality_median_residual}
   \end{minipage}
   \caption{Median S-DeepONet case on Rel. $L_2$ error, qualitative comparison. Results are shown at a selected timestep that illustrates typical model performance. Residuals are computed as the difference between each model's prediction and the ground truth (GT) solution.}
   \label{ldc_quality_median}
\end{figure}

\begin{figure}[h!]
   \centering
   \begin{minipage}{0.49\textwidth}
       \centering
       \includegraphics[width=\textwidth]{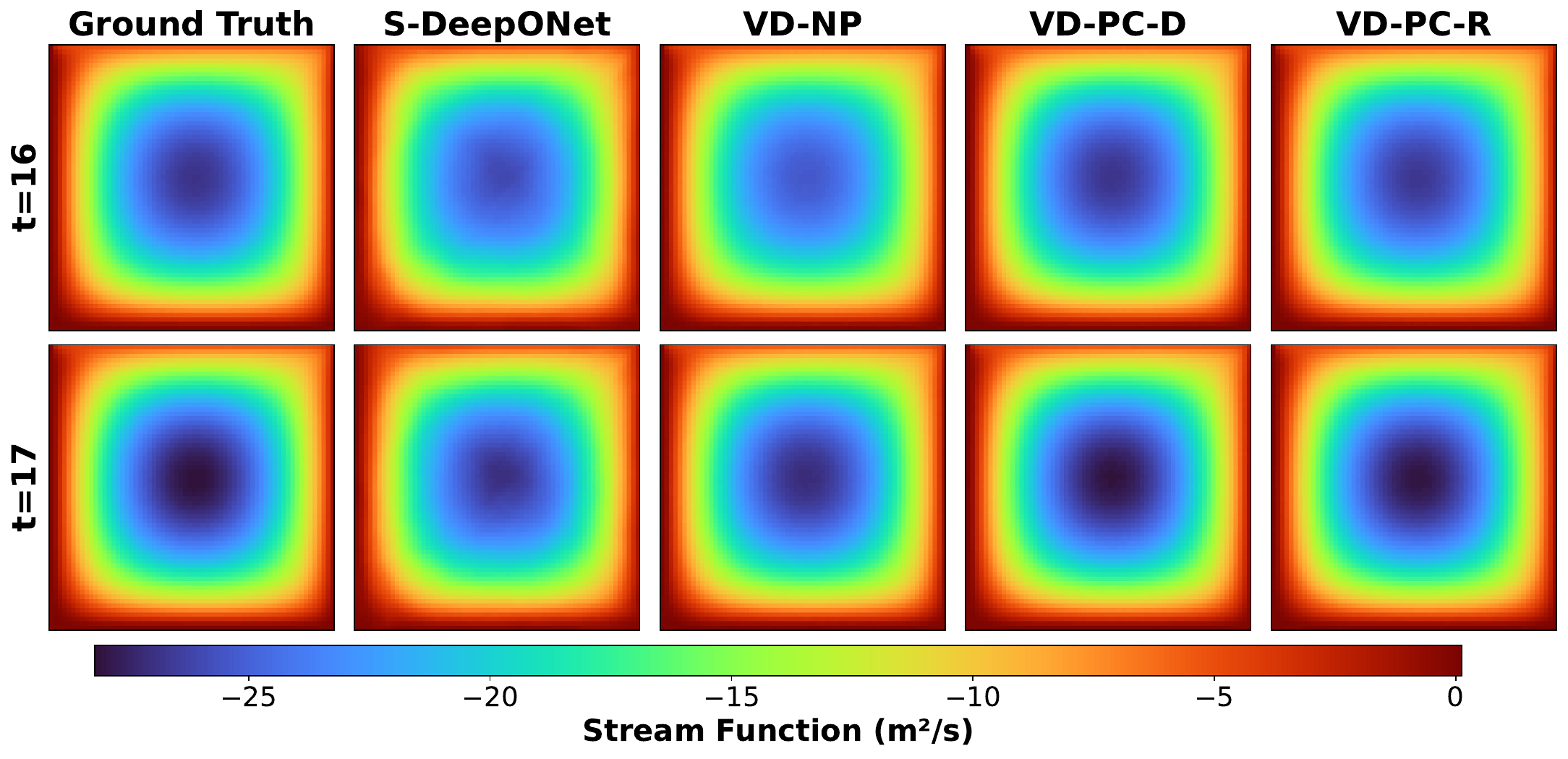}
       \subcaption{Stream function comparison}
       \label{ldc_quality_worst_full_field}
   \end{minipage}
   \begin{minipage}{0.49\textwidth}
       \centering
       \includegraphics[width=\textwidth]{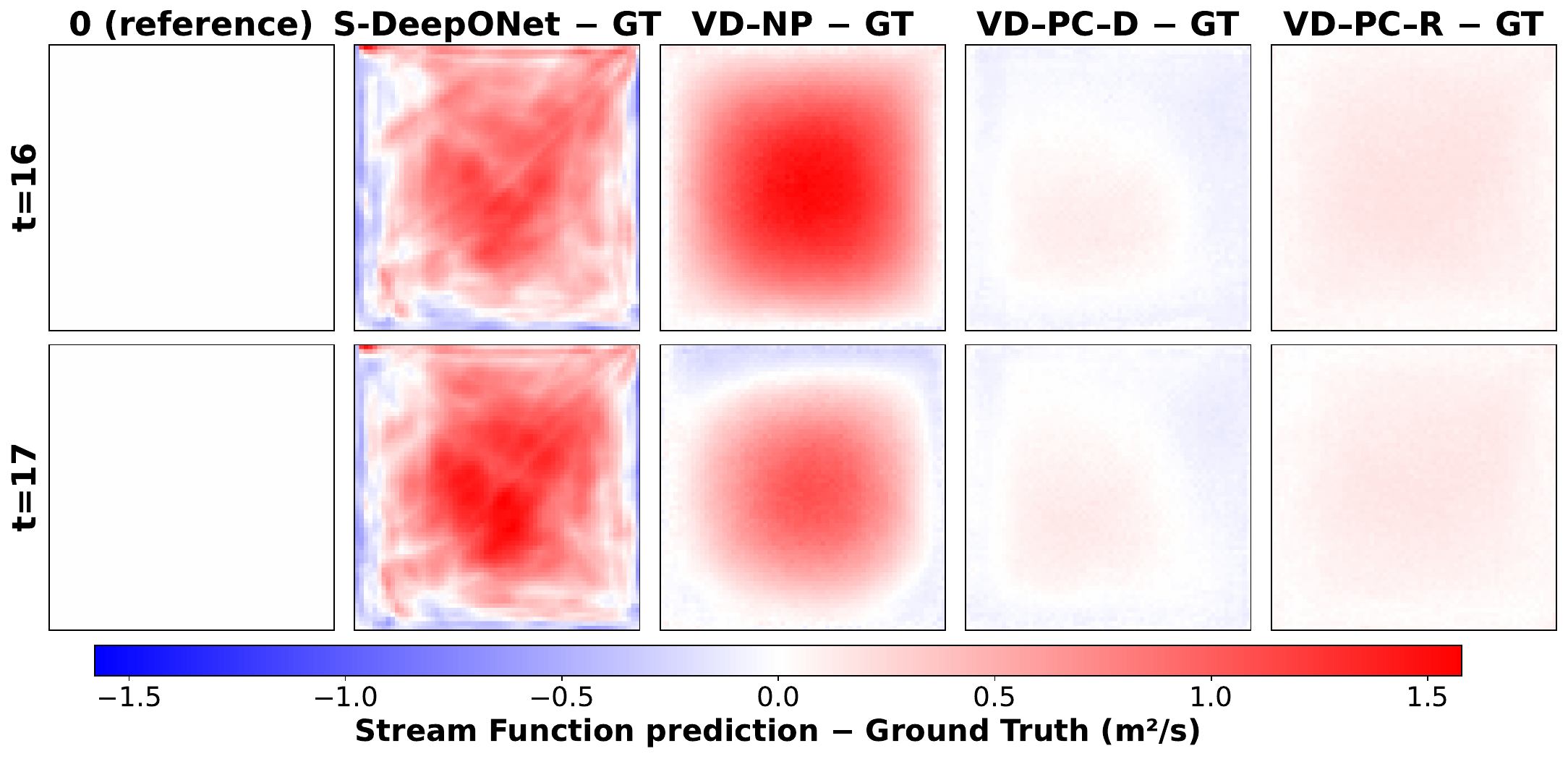}
       \subcaption{Residual comparison}
       \label{ldc_quality_worst_residual}
   \end{minipage}
   \caption{Worst S-DeepONet case on Rel. $L_2$ error, qualitative comparison. Residuals are computed as the difference between each model's prediction and the ground truth (GT) solution.}
   \label{ldc_quality_worst}
\end{figure}

For qualitative analysis, we present predictions for each models (S-DeepONet, VD-NP, VD-PC-D, and VD-PC-R) with the true stream function video. \fref{ldc_preds} illustrates the overall idea of what each model outputs and \fref{ldc_quality_best}, \fref{ldc_quality_median}, \fref{ldc_quality_worst} demonstrates the video quality by checking both the predicted stream function field and their residual field. For \fref{ldc_preds}, timeframes (t = 1, 2, 3, 4, 5, 6) were taken where the flow field is temporally most evolving in true solution field, to show how each model performs on the most challenging phases.

\fref{ldc_preds} juxtaposes ground truth stream function fields with the four surrogates at six earliest frames $(t=1 \text{ to } 6)$ and the final frame (t=40). The standalone S-DeepONet captures only the gross circulation pattern while systematically blurring corner vortices and smearing shear layers, resulting in poor qualitative fidelity. The standalone video diffusion (VD‐NP) recovers the overall topology more faithfully, yet it exhibits high errors in later frames (further illustrated in
\fref{ldc_quality_best}--\ref{ldc_quality_worst}) and introduces grainy speckle artifacts. By contrast, the hybrid S-DeepONet-conditioned diffusion models (VD-PC-D and VD-PC-R) reconstructs both the primary vortex and the fine peripheral features with nearly indistinguishable color bands with the true video. This improvement suggests that leveraging the S‐DeepONet
prior not only reinforces coherent flow structures in all timesteps but also suppresses the
speckle artifacts seen in the baselines. These qualitative observations align with the one‐order‐of‐magnitude error reductions documented earlier, underscoring the benefits of coupling S-DeepONet with video diffusion corrector.

\begin{comment}
\begin{figure}[h!]
    \centering
    \includegraphics[width=0.8\textwidth]{ldc_quality_check.pdf}
    
    \vspace{0.2cm}
    
    \includegraphics[width=0.8\textwidth]{ldc_quality_check_residual.pdf}
    \caption{Qualitative comparison of stream function of best case for S-DeepONet prediction.}
    \label{ldc_quality}
\end{figure}  
\end{comment}

To capture clearer difference between each models, we visualize three additional samples in \fref{ldc_quality_best}, \fref{ldc_quality_median}, \fref{ldc_quality_worst}, corresponding to the best, median, and worst SDON cases predicted by the standalone S-DeepONet respectively. Each figure contains two panels: (a) the predicted stream‑function field and (b) the residual (\(\hat s_{\text{pred}} -s_{\text{true}}\)). Although the standalone video diffusion model (VD‐NP) appears visually accurate at earlier timesteps (\fref{ldc_preds}), its errors peak in later frames; for instance, at $t=35-36$ in \fref{ldc_quality_best_residual}, VD‐NP exhibits the most prominent residuals. The same pattern recurs in \fref{ldc_quality_median_residual} and \fref{ldc_quality_worst_residual}, confirming that VD‐NP struggles with longer‐term evolution, providing only the input function $\bm{a}$ as a guidance is too weak for generating temporally consistent video. Meanwhile, the standalone S-DeepONet predictions reveal inconsistent circulation shapes and smeared spatial gradients, which stand out even more clearly in the residual panels. By contrast, the two S‐DeepONet‐conditioned diffusions (VD‐PC‐D and VD‐PC‐R) maintain accurate flow structures at late timesteps as well, showing predominantly white (near‐zero) residual fields that affirm their spatiotemporal robustness, consistent with the quantitative errors reported earlier. Detailed residual field results for the best case, median case, and the worst case is reported at \fref{appendix_ldc_quality_best}, \fref{appendix_ldc_quality_median}, and \fref{appendix_ldc_quality_worst}.

\subsection{Predicting dogbone plasticity}

\paragraph{Quantitative result analysis}

\begin{table}[h!]
\centering
\caption{Overall statistics of error metrics for the von Mises stress field in the elasto-plastic dogbone specimen benchmark.}
\begin{tabularx}{\textwidth}{ 
  >{\centering\arraybackslash}X 
  >{\centering\arraybackslash}X 
}
\begin{tabular}{c|c|c|c|c}
\multicolumn{5}{c}{\textbf{Elasto-plastic tensile simulation of dogbone sample}} \\ \hline
\textbf{}  & \textbf{S-DeepONet} & \textbf{VD-NP} & \textbf{VD-PC-D} & \textbf{VD-PC-R} \\ \hline
Mean Rel. $L_2$      & 4.43\%       & 123\%       & 3.17\%      & 2.94\%    \\ 
Mean RMAE            & 4.11\%       & 133\%       & 3.11\%      & 2.95\%    \\
Mean MAE             & 1.67 MPa     & 46.7 MPa    & 1.33 MPa    & 1.20 MPa
\end{tabular}
\end{tabularx}
\label{tab:dogbone_error_overall}
\end{table}

We now apply the same framework to the von Mises stress‐field prediction for the dogbone tensile‐plasticity benchmark. \tref{tab:dogbone_error_overall} gives the mean errors defined in \eref{eval_metrics} to show overall model performance. Additional metric the MAE was calculated. The need of this metric is later discussed in this section. All metrics, the mean Rel. $L_2$ error, the mean RMAE, and the mean MAE agrees on utilizing the S-DeepONet information to the video diffusion model and setting residual field as a target (VD-PC-R) provides the best result consistent with the previous benchmark.

\tref{tab:l2_plasticity_percentile} and \tref{tab:mae_plasticity_percentile} summarize key statistics for the relative $L_2$ error and the relative mean‐absolute error (RMAE). Compared to the lid‐driven‐cavity flow, this dataset proves more challenging, resulting in larger worst‐case errors for all methods.

\begin{table}[h!]
\centering
\caption{Percentile statistics of the relative $L_2$ error (\%) for the von Mises stress field in the tensile plasticity on dogbone sample benchmark.}
\begin{tabularx}{\textwidth}{ 
  >{\centering\arraybackslash}X 
  >{\centering\arraybackslash}X 
}
\begin{tabular}{c|c|c|c|c|c}
\multicolumn{6}{c}{\textbf{Rel. $L_2$ error by each percentile}} \\ \hline
\textbf{}  & \textbf{Best case} & $\mathbf{25^{th}}$ & $\mathbf{50^{th}}$ & $\mathbf{75^{th}}$ & \textbf{Worst case} \\ \hline
S-DeepONet                       & 1.01\%       & 2.70\%      & 3.73\%      & 5.60\%     & 18.1\%    \\ 
VD-NP                  & 22.2\%       & 74.1\%      & 100\%      & 146\%     & 670\%    \\ 
VD-PC-D  & 0.880\%      & 1.84\%     & 2.53\%     & 3.80\%     & 16.2\%    \\
\textbf{VD-PC-R}  & 0.658\%      & 1.65\%     & 2.34\%     & 3.52\%     & 15.3\%  
\end{tabular}
\end{tabularx}
\label{tab:l2_plasticity_percentile}
\end{table}

\begin{table}[h!]
\centering
\caption{Percentile statistics of the relative mean absolute error (\%) for the von Mises stress field in the tensile plasticity on dogbone sample benchmark.}
\begin{tabularx}{\textwidth}{ 
  >{\centering\arraybackslash}X 
  >{\centering\arraybackslash}X 
}
\begin{tabular}{c|c|c|c|c|c}
\multicolumn{6}{c}{\textbf{RMAE by each percentile}} \\ \hline
\textbf{}  & \textbf{Best case} & $\mathbf{25^{th}}$ & $\mathbf{50^{th}}$ & $\mathbf{75^{th}}$ & \textbf{Worst case} \\ \hline
S-DeepONet                       & 0.806\%       & 2.19\%      & 3.32\%      & 5.26\%     & 17.4\%    \\ 
VD-NP                  & 23.1\%       & 77.3\%      & 110\%      & 161\%     & 671\%    \\ 
VD-PC-D  & 0.756\%      & 1.69\%     & 2.42\%     & 3.79\%     & 16.4\%    \\
\textbf{VD-PC-R}  & 0.594\%      & 1.57\%     & 2.32\%     & 3.61\%     & 16.2\%
\end{tabular}
\end{tabularx}
\label{tab:rmae_plasticity_percentile}
\end{table}

\begin{figure}[h!] 
    \centering
    \begin{minipage}{0.48\textwidth}
        \centering
        \includegraphics[width=\textwidth]{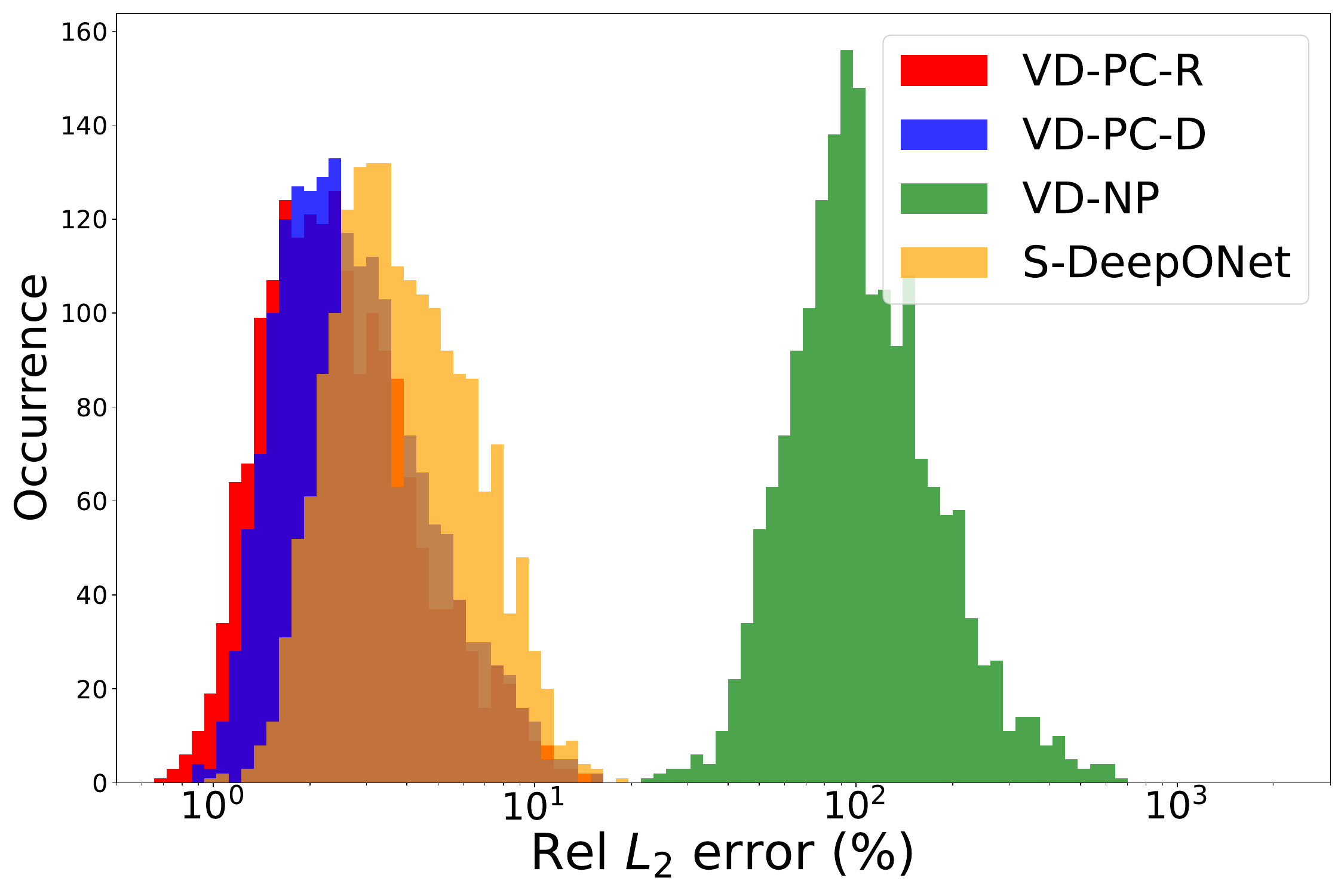}
        \caption{Histogram of Rel. $L_2$ error for plastic deformation on dogbone example.}
        \label{fig:dogbone_l2_hist}
    \end{minipage}
    \hfill
    \begin{minipage}{0.48\textwidth}
        \centering
        \includegraphics[width=\textwidth]{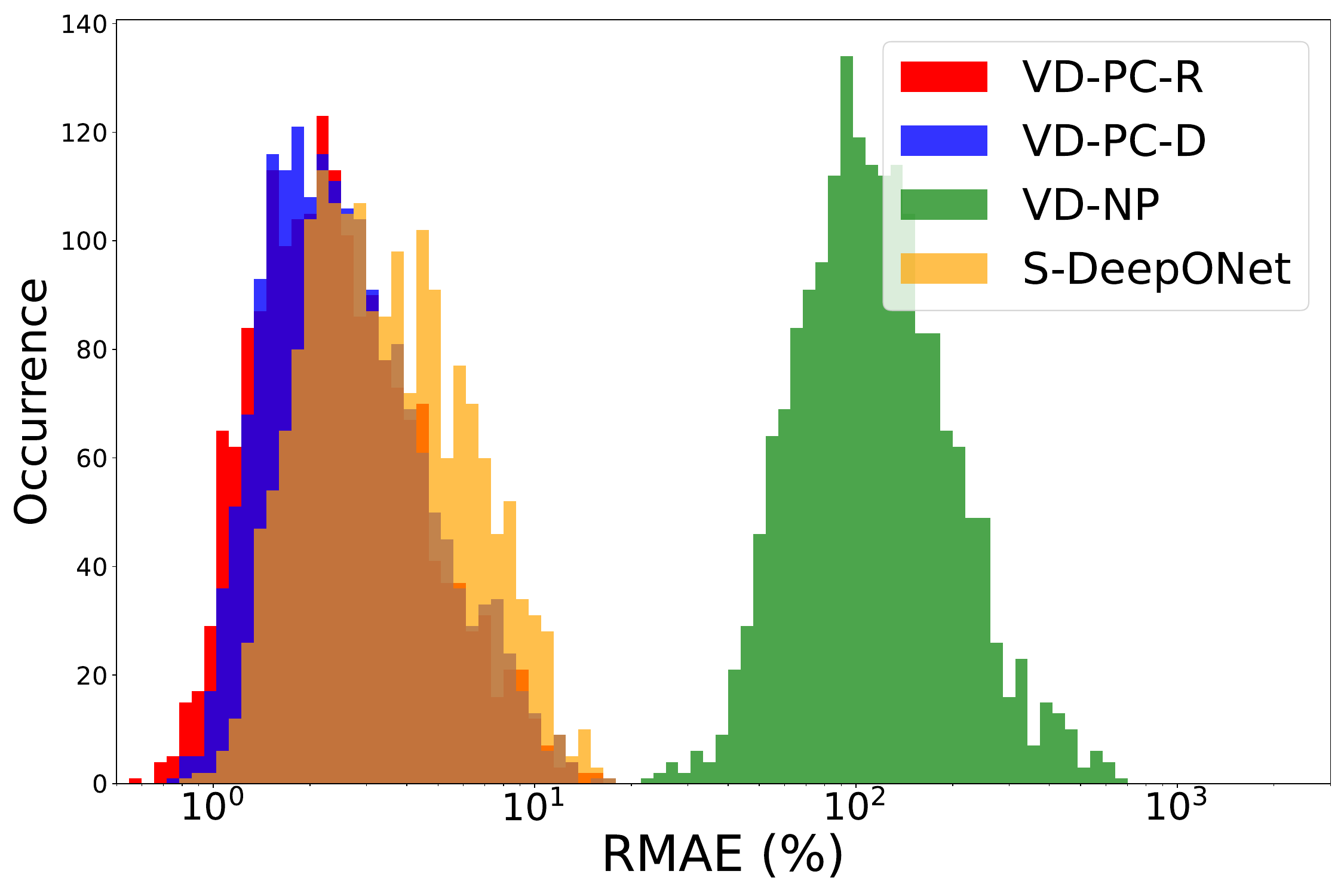}
        \caption{Histogram of RMAE for plastic deformation on dogbone example.}
        \label{fig:dogbone_rmae_hist}
    \end{minipage}
\end{figure}

A key noticeable fact in this example is that the standalone video diffusion model performed significantly poor: 100\% and 110\% relative $L_2$ error and RMAE for the median cases, and can spike to 670\% in the worst cases. Such huge error stems from how we defined the relative errors \eref{eval_metrics}. Due to the presence of near‐zero ground‐truth frames in this certain dataset, even small absolute discrepancies cause the relative error to blow up. Nonetheless, the high error rates clearly indicate that VD‐NP fails to faithfully map the time‐dependent load function~$\bm{a}$ to the resulting plastic‐stress field.

To mitigate these artifacts, we introduce one more metric: namely the mean absolute error (MAE) with dimensions of MPa,
\begin{equation}
\begin{aligned}
    \text{Mean absolute error}
    \;&=\;
    \bigl\lVert s_{\text{True}} - s_{\text{Pred}} \bigr\rVert_{1}, \quad \text{where} \\
    \|A\|_1&=\frac{1}{T H W} \sum_{t=1}^T \sum_{h=1}^H \sum_{w=1}^W\left|A_{t, h, w}\right| .
\end{aligned}
\label{eq:mae_definition}
\end{equation}

\begin{table}[h!]
\centering
\caption{MAE statistics for the von Mises stress field in the tensile plasticity on dogbone sample benchmark.}
\begin{tabularx}{\textwidth}{ 
  >{\centering\arraybackslash}X 
  >{\centering\arraybackslash}X 
}
\begin{tabular}{c|c|c|c|c|c}
\multicolumn{6}{c}{\textbf{MAE by each percentile}} \\ \hline
\textbf{[MPa]}  & \textbf{Best case} & $\mathbf{25^{th}}$ & $\mathbf{50^{th}}$ & $\mathbf{75^{th}}$ & \textbf{Worst case} \\ \hline
S-DeepONet        & 0.950      & 1.41       & 1.62       & 1.88       & 3.27    \\ 
VD-NP             & 17.6       & 40.1       & 45.9       & 52.5       & 90.8    \\ 
VD-PC-D           & 0.804      & 1.15       & 1.29       & 1.46       & 2.85    \\
\textbf{VD-PC-R}  & 0.614      & 1.01       & 1.16       & 1.33       & 2.87
\end{tabular}
\end{tabularx}
\label{tab:mae_plasticity_percentile}
\end{table}

\begin{figure}[h!] 
    \centering
         \includegraphics[width=0.5\textwidth]{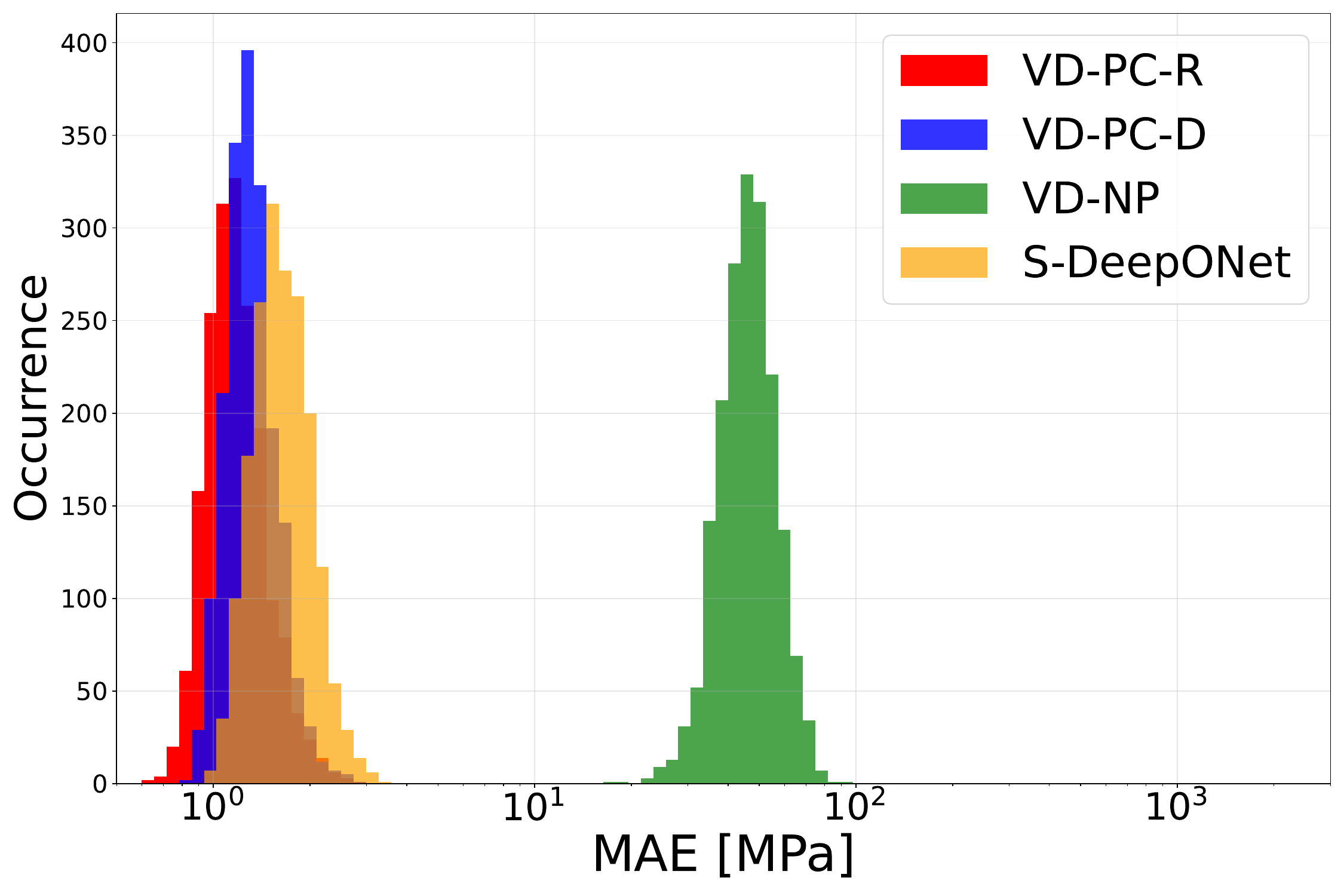}
    \caption{Histogram of the Mean absolute error for plastic deformation on dogbone example.}
    \label{fig:dogbone_mae_hist}
\end{figure}

\tref{tab:mae_plasticity_percentile} and \fref{fig:dogbone_mae_hist} confirm that VD‐NP again exhibits relatively large MAE (up to 90.8 MPa), whereas the two hybrid models (VD‐PC‐D and VD‐PC‐R) reduce the median MAE to roughly 1 MPa in the median cases, well below the standalone S‐DeepONet’s 1.62 MPa. This finding echoes the earlier conclusions from the previous section, reaffirming that S‐DeepONet‐guided diffusion not only achieves higher spatial accuracy but also exhibits greater robustness across different timeframes. Overall, these results demonstrate that conditioning the diffusion model with the neural‐operator prior enables it to closely track the true solution field, even under significant plastic‐strain gradients. \fref{fig:dogbone_mae_hist} shows MAE histograms with x-axis log scale, proving VD-PC-R generates solution field that are the closest to the ground truth.

Aside from huge relative error for VD-NP, the relative error metrics agrees with the trend. As shown in \ref{tab:l2_plasticity_percentile}, and \ref{tab:rmae_plasticity_percentile} the worst‐case relative $L_2$ error and RMAE for the two hybrid approaches (VD‐PC‐D and VD‐PC‐R) hover around 15 - 16 \%, enhancement compared to the standalone S-DeepONet by 15\%. From the median perspective, S‐DeepONet attains 3.73\% $L_2$ error, which the hybrid models reduce to 2.34\% an enhancement by 37\%. It goes same with RMAE with S-DeepONet’s 3.32\% dropping to 2.32\%, a relative gain of 30\%. These benefits are consistent across the best case, 25th and 75th percentiles, indicating that the hybrid corrects a wide range of typical cases rather than only the easiest scenarios. The error histograms in \fref{fig:dogbone_l2_hist}, and \fref{fig:dogbone_rmae_hist} reinforce these observations: both VD‐PC‐D and VD‐PC‐R shift most samples into the lower‐error bins, while VD‐NP’s distribution stretches so far to the right that a log scale was needed on the \(x\)‐axis. This stark contrast again underscores that merely providing the load function $\bm{a}$ does not suffice for stable, accurate stress‐field generation whereas the additional prior from S‐DeepONet significantly mitigates outliers and achieves more reliable predictions across the entire test set.

\paragraph{Qualitative result analysis}

\begin{figure}[h!]
    \centering
         \includegraphics[width=1.0\textwidth]{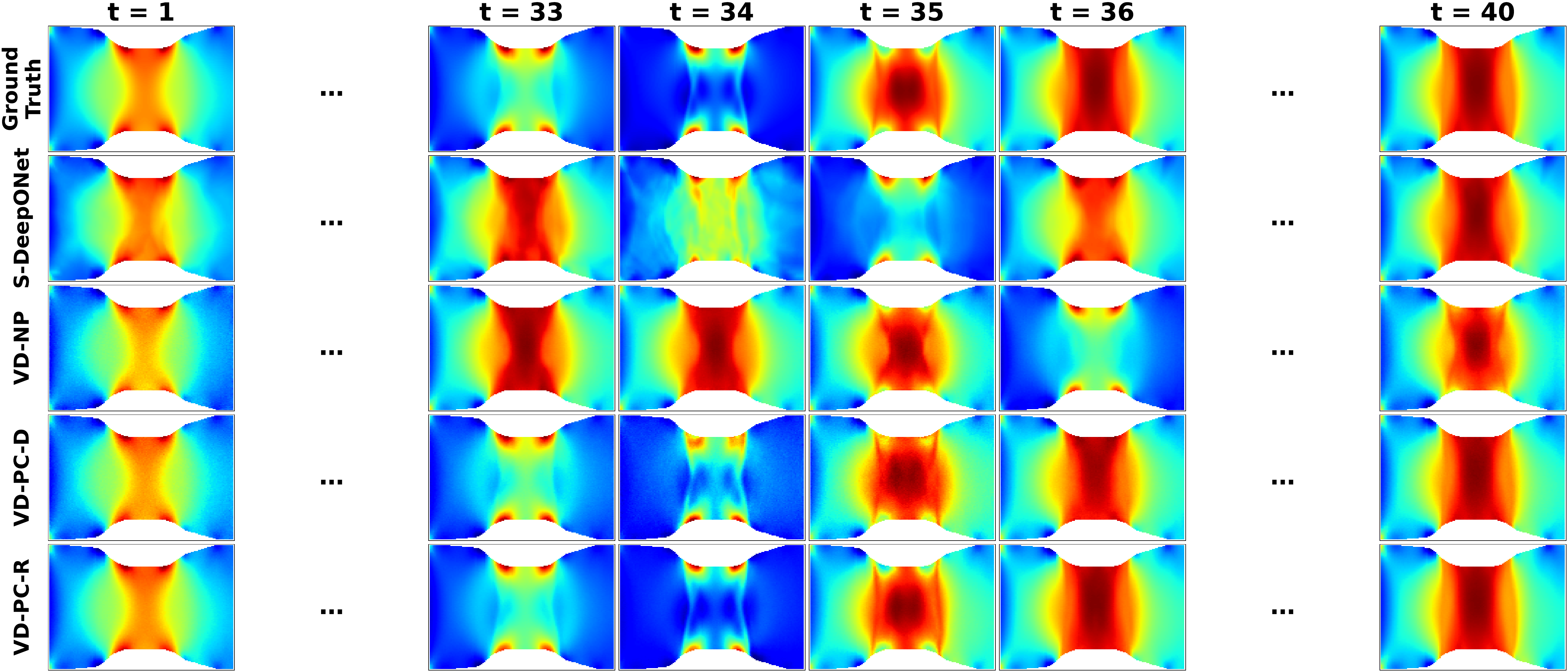}
    \caption{Temporal evolution of von Mises stress field predictions across critical timesteps during transient loading. Each row shows predictions from different approaches: ground truth finite element results, baseline S-DeepONet, pure diffusion model, and the proposed hybrid conditioned diffusion methods (both diffusion-conditioned and residual-conditioned variants). The selected timesteps (t = 33, 34, 35, 36) capture the progression from initial loading through peak stress evolution to final equilibrium, highlighting how each model handles the most challenging transient phases where stress field gradients are evident and temporal variations are most pronounced.}
    \label{dogbone_preds}
\end{figure}

Inheriting the previous example, \fref{dogbone_preds} compares the predicted stress‐field videos at four
transient timesteps (33–36), when the material experiences its most significant changes in its ground truth data. The standalone S-DeepONet yields a spatially smeared solution with inconsistent internal‐stress shapes, where the VD-NP model maintains better shape consistency yet fails to track the ground‐truth
temporal evolution, but it fails to temporally follow the ground truth data. This is one reason for the large errors observed in the quantitative analysis. By contrast, the two prior‐conditioned diffusion models (VD‐PC‐D and VD‐PC‐R) achieve higher fidelity both spatially and temporally, maintaining stress patterns close to the reference fields at each timestep. In this example, VD‐PC‐R further reduces pixelation artifact relative to VD‐PC‐D, illustrating how residual‐learning can mitigate fine‐scale artifacts by learning a reduced problem.

\begin{figure}[h!]
   \centering
   \begin{minipage}{0.49\textwidth}
       \centering
       \includegraphics[width=\textwidth]{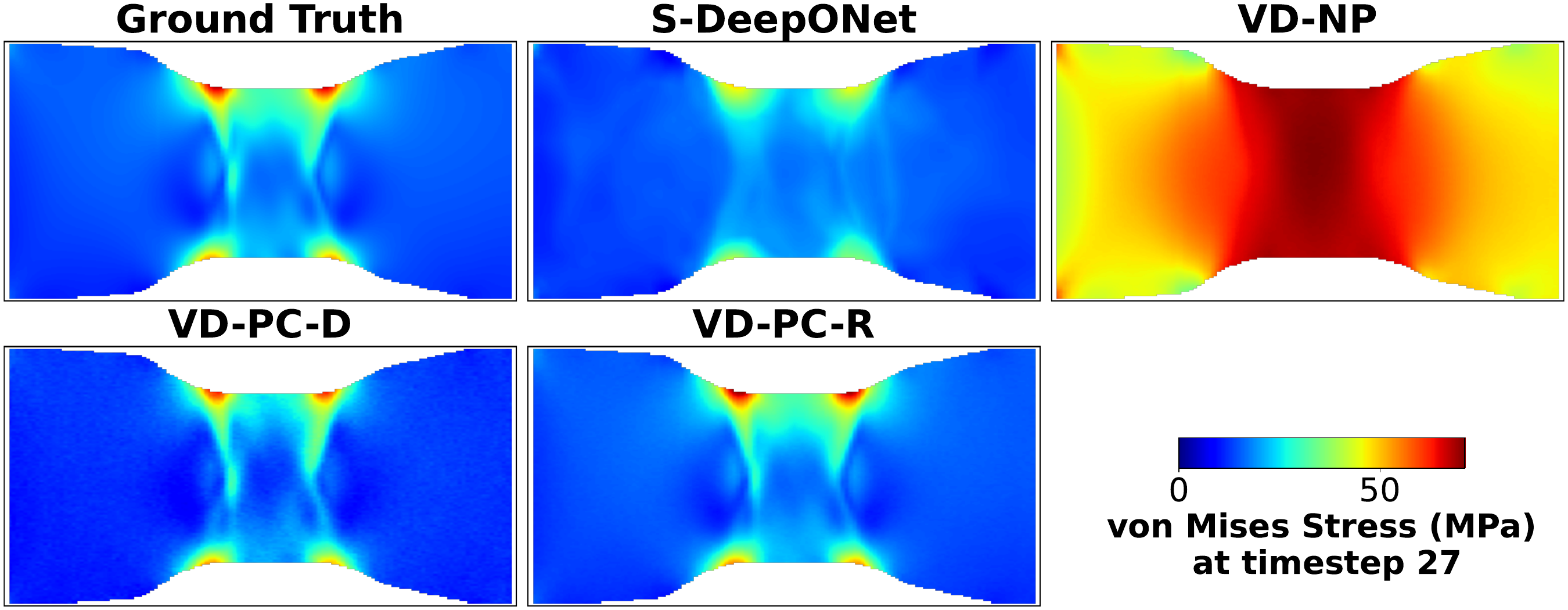}
       \subcaption{Stress field comparison}
       \label{dogbone_quality_best_full_field}
   \end{minipage}
   \begin{minipage}{0.49\textwidth}
       \centering
       \includegraphics[width=\textwidth]{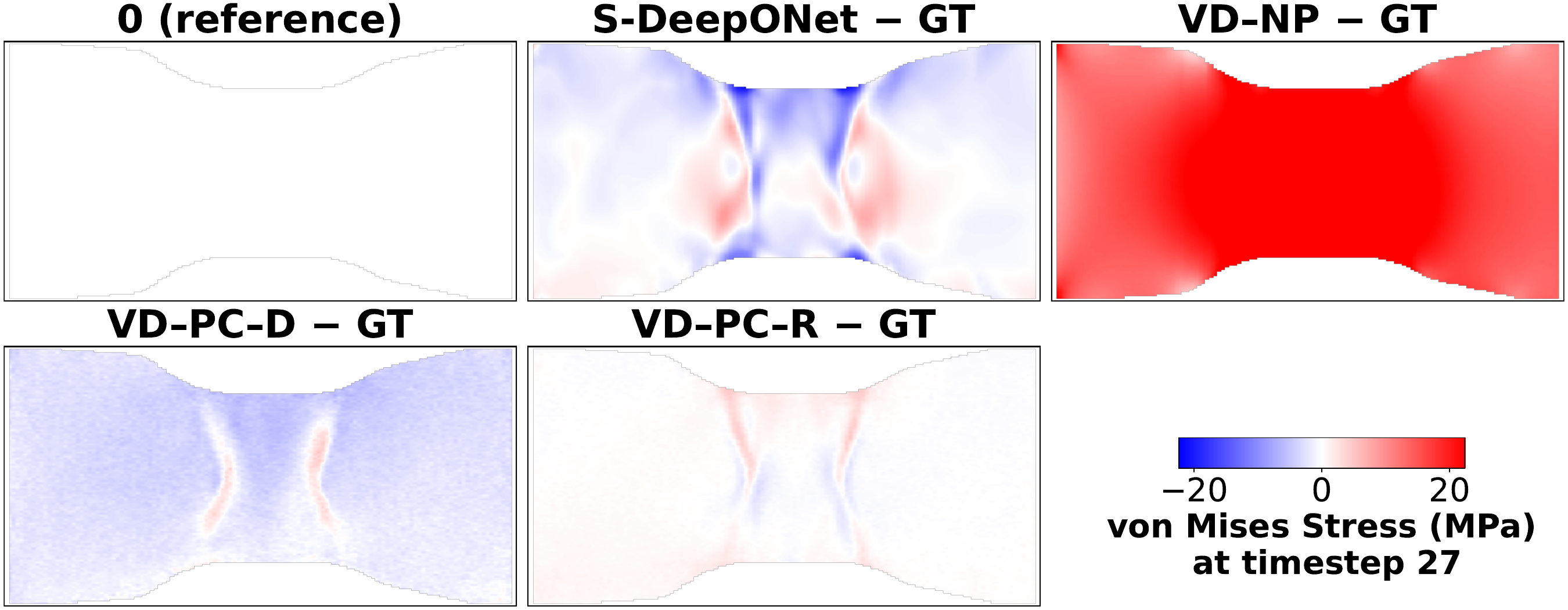}
       \subcaption{Residual comparison}
       \label{dogbone_quality_best_residual}
   \end{minipage}
   \caption{Best pereforming S-DeepONet case on MAE, qualitative comparison. Residuals are computed as the difference between each model's prediction and the ground truth (GT) solution.}
   \label{dogbone_quality_best}
\end{figure}

\begin{figure}[h!]
   \centering
   \begin{minipage}{0.49\textwidth}
       \centering
       \includegraphics[width=\textwidth]{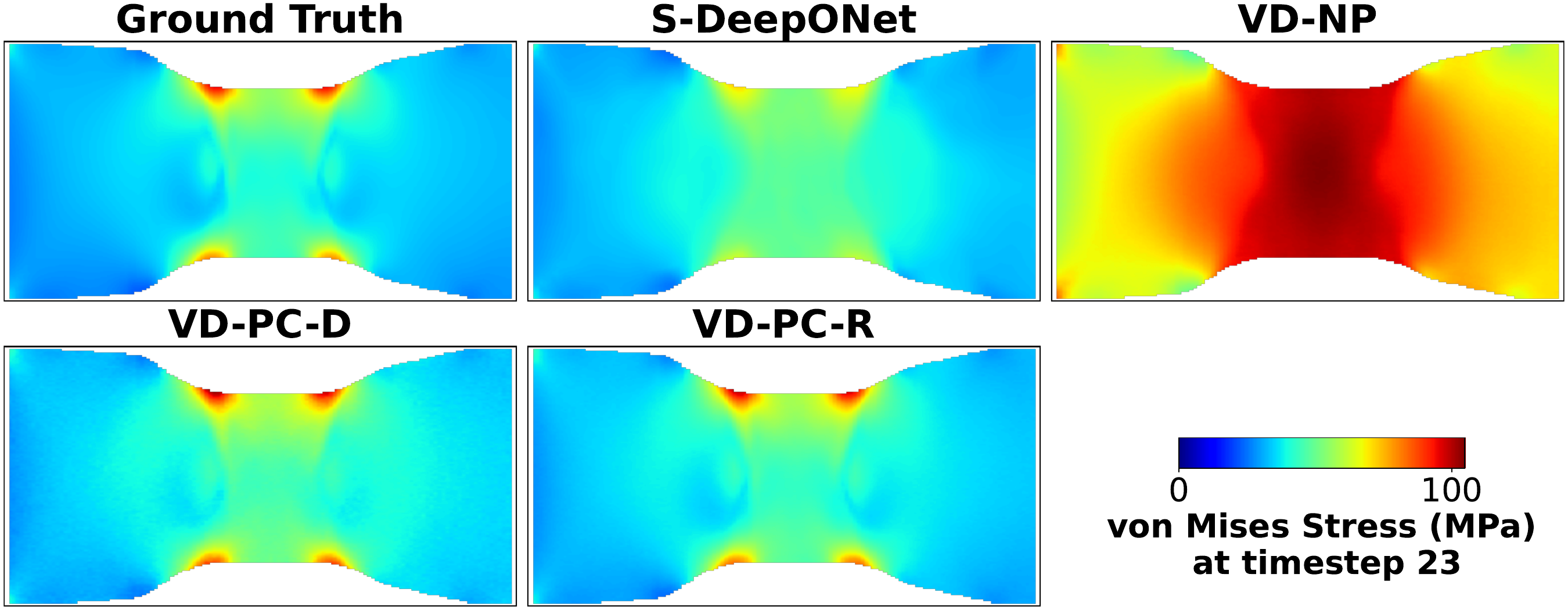}
       \subcaption{Stress field comparison}
       \label{dogbone_quality_median_full_field}
   \end{minipage}
   \begin{minipage}{0.49\textwidth}
       \centering
       \includegraphics[width=\textwidth]{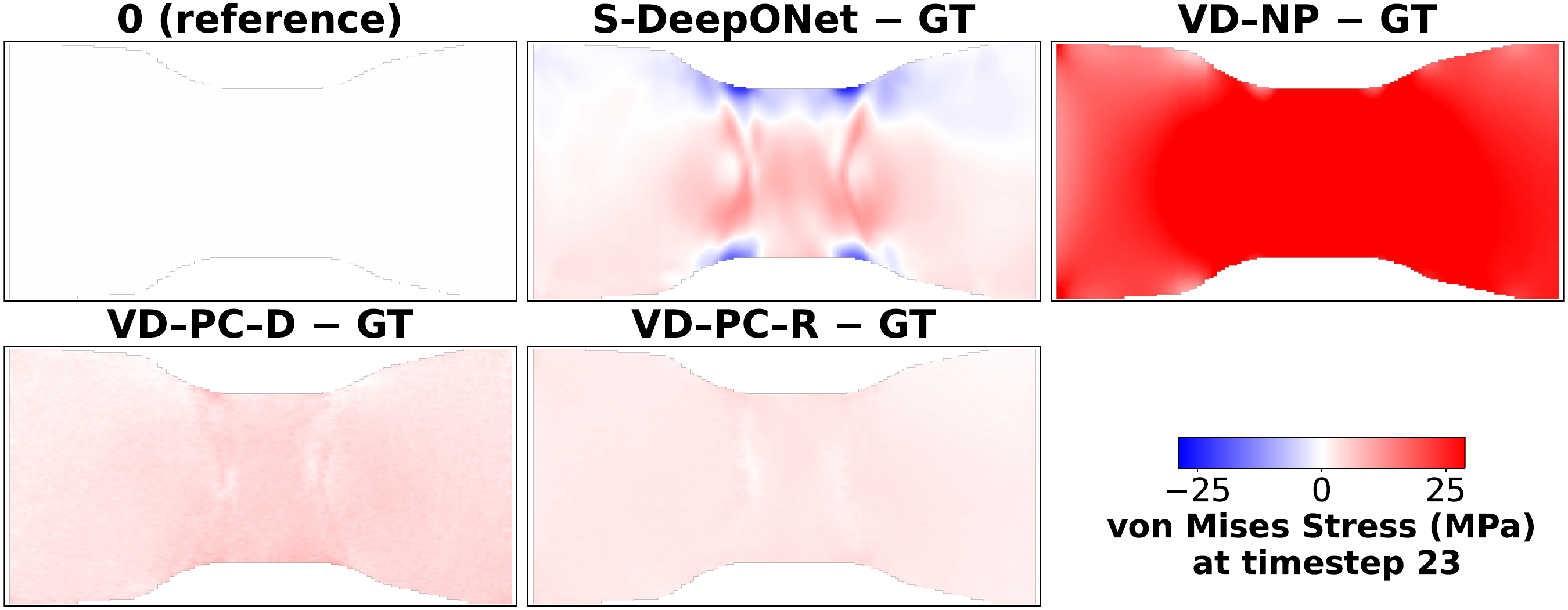}
       \subcaption{Residual comparison}
       \label{dogbone_quality_median_residual}
   \end{minipage}
   \caption{Median S-DeepONet case on MAE, qualitative comparison. Residuals are computed as the difference between each model's prediction and the ground truth (GT) solution.}
   \label{dogbone_quality_median}
\end{figure}

\begin{figure}[h!]
   \centering
   \begin{minipage}{0.49\textwidth}
       \centering
       \includegraphics[width=\textwidth]{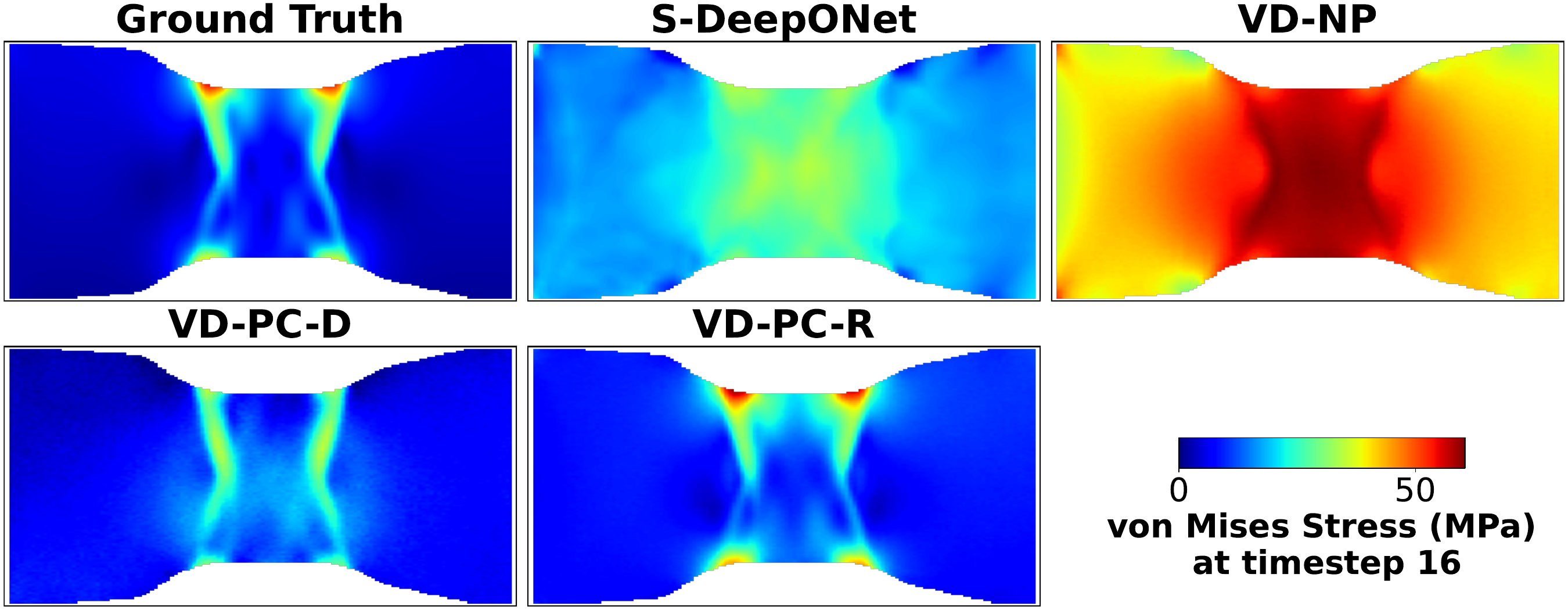}
       \subcaption{Stress field comparison}
       \label{dogbone_quality_worst_full_field}
   \end{minipage}
   \begin{minipage}{0.49\textwidth}
       \centering
       \includegraphics[width=\textwidth]{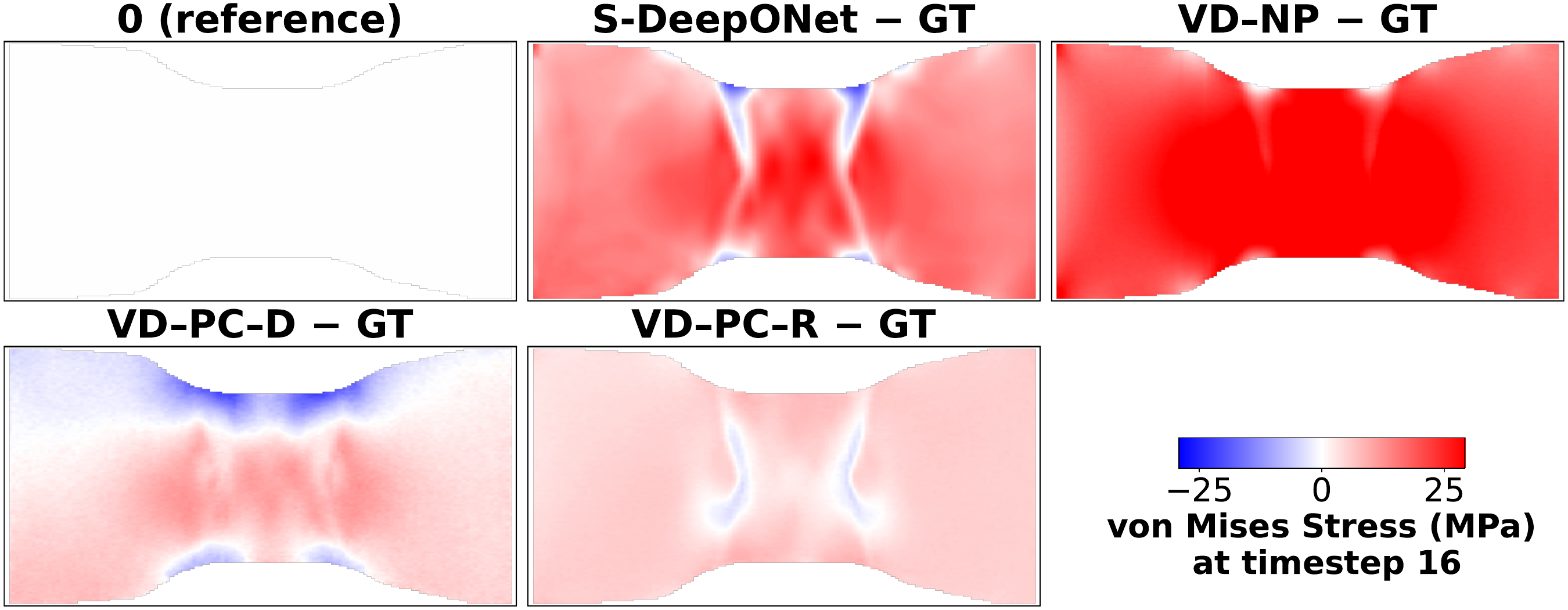}
       \subcaption{Residual comparison}
       \label{dogbone_quality_worst_residual}
   \end{minipage}
   \caption{Worst S-DeepONet case on MAE, qualitative comparison. Residuals are computed as the difference between each model's prediction and the ground truth (GT) solution.}
   \label{dogbone_quality_worst}
\end{figure}

To probe a broader range of scenarios, we provide three additional examples: \fref{dogbone_quality_best}, \fref{dogbone_quality_median}, and \fref{dogbone_quality_worst} corresponding to the best, median, and worst S‐DeepONet predictions in terms of MAE respectively. Each figure comprises two panels: (a) the stress‐field comparison and (b) the residual (\(\hat s_{\text{pred}}-s_{\text{true}}\)). For each cases, a certain timestep was taken to visualize stress field where the model differences can be most visually identified. In every case,
VD‐NP drifts even further from the reference solution, with its residual maps overshooting the color scale and appearing predominantly red. Aside from VD-NP, the same qualitative trends observed in the cavity flow problem recur: S‐DeepONet alone captures the broad stress distribution but suffers from irregularities and smearing, while the two hybrid (S‐DeepONet‐conditioned) diffusions remain closer to the ground truth and exhibit smaller residuals. In particular, conditioning diffusion on the operator prior removes artifacts and simultaneously sharpens the stress hotspots at the fillet corners, yielding a map that is closest to the ground truth. Agreeing with the quantitative analysis in the previous section, VD-PC-R outperforms VD-PC-D in terms of residual magnitudes, reinforcing the idea that residual‐based conditioning lets the diffusion stage focus on fine‐scale corrections. Taken together, these results confirm that even for highly nonlinear elasto‐plastic processes, providing a physics‐aware coarse prior significantly stabilizes the generative trajectory and achieves systematically higher accuracy. Extra timesteps for each cases in \fref{dogbone_quality_best}, \fref{dogbone_quality_median}, and \fref{dogbone_quality_worst} are further illustrated in \fref{appendix_dogbone_quality_best}, \fref{appendix_dogbone_quality_median}, and \fref{appendix_dogbone_quality_worst}.

\subsection{Runtime comparison}
\label{sec:runtime}

\begin{figure}[h!]
    \centering
    \begin{minipage}{0.48\textwidth}
        \centering
        \includegraphics[width=\textwidth]{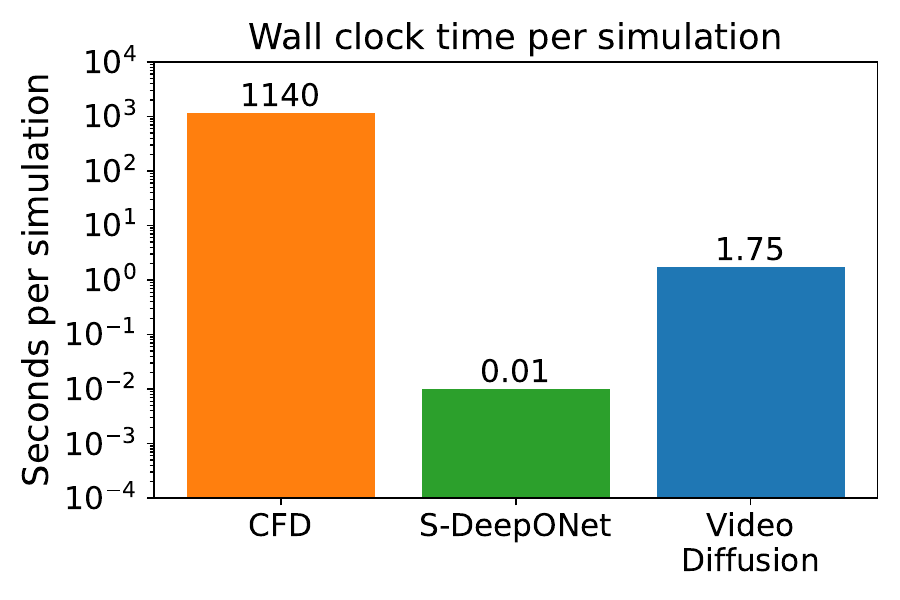}
        \caption{Wall clock time per lid-driven-cavity case:
                 high-fidelity CFD solver vs.\ S-DeepONet (SDON) and the three
                 diffusion variants, which have nearly identical sampling
                 times and are therefore grouped together.}
        \label{fig:ldc_runtime}
    \end{minipage}
    \hfill
    \begin{minipage}{0.48\textwidth}
        \centering
        \includegraphics[width=\textwidth]{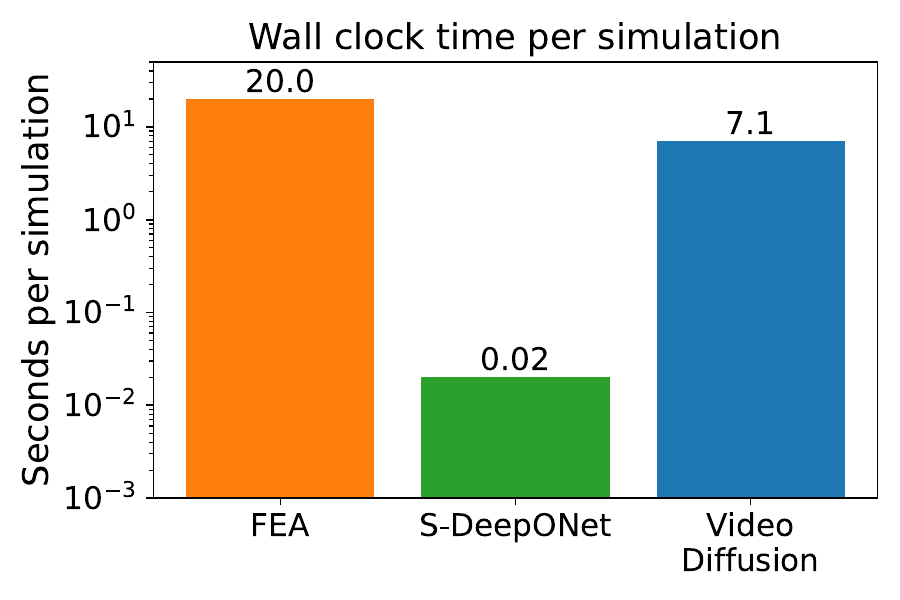}
        \caption{Wall clock time per dogbone plasticity case:
                 FEA vs.\ SDON and the grouped diffusion
                 surrogates.}
        \label{fig:dogbone_runtime}
    \end{minipage}
\end{figure}

\fref{fig:ldc_runtime} and \fref{fig:dogbone_runtime} compare the wall-clock time required to generate a single solution video by (i) the high-fidelity solver, (ii) the stand-alone S-DeepONet, and (iii) the family of video-diffusion models on average.\footnote{VD-NP, VD-PC-D, and VD-PC-R share the same 3-D U-Net backbone and thus have virtually identical sampling times; they are plotted as a single bar.} For the lid-driven-cavity benchmark, a turbulent CFD run takes $\approx\!1\,140\,$s ($\sim\!19\,$min), whereas SDON delivers a prediction in $1.0\times10^{-2}\,$s and the diffusion corrector in $\approx\!1.7\,$s three to five orders of magnitude faster than the high-fidelity solver while remaining two orders slower than the operator alone. In the plasticity benchmark, the FEA solver is shorter
($\approx\!20\,$s) thanks to optimized implicit integration \cite{Abaqus2024}, yet the same hierarchy persists: SDON at $2.0\times10^{-2}\,$s, diffusion at $\approx\!7\,$s, and the FEA baseline an order of magnitude slower. Overall, these results show that S-DeepONet provides instantaneous coarse predictions, while the prior-conditioned diffusion adds $\mathcal{O}(10)$ s of compute to recover high-resolution details—still a two–to–three-order-of-magnitude speed-up over conventional solvers and therefore practical for design loops or real-time monitoring. Further runtime reductions via model-distillation techniques (see \ref{sec:conc}) could close the remaining gap without sacrificing accuracy.

\section{Conclusions and future work}
\label{sec:conc}

We presented a hybrid surrogate framework that combines a Sequential Deep Operator Network (S‐DeepONet) with a conditional video‐diffusion decoder to predict spatio‐temporal solution fields governed by nonlinear partial
differential equations. In both the lid‐driven‐cavity flow and the dogbone‐plasticity benchmarks, our approach outperformed standalone models (S‐DeepONet alone or diffusion alone) by a notable margin, consistently reducing errors and improving visual fidelity. These results underscore the value of a physics‐aware prior, which constrains the diffusion process to refine primarily the high‐frequency and localized structures that the operator network alone struggles to capture.

We further demonstrated two modes of diffusion conditioning: generating the full‐field solution directly, or predicting only the pointwise residual relative to the S‐DeepONet prior.  Across all experiments, residual‐based diffusion (VD‐PC‐R) yielded the most accurate and stable predictions, revealing that learning the discrepancy focuses the model capacity on fine‐scale corrections.  This not only boosted accuracy but also mitigated speckle or pixelation artifacts that can arise when diffusion must learn the entire solution from scratch.

For future work, we envision several directions:

\begin{itemize}
    \item \emph{Extended physics domains.} While we verified the method on fluid flow and plasticity, similar operator‐plus‐diffusion architectures may benefit other applications, including multiphase flows, fracture mechanics, or even thermo‐mechanical coupling.

    \item \emph{Diffusion model distillations.} While diffusion model sampling can be done in the order of 10 seconds per simulation which is significantly faster than FEA/CFD simulations, it still is slower compared to S-DeepONet inference time which is in the order of $10^{-2}$ seconds per case. Use of progressive distillation \cite{salimans2022progressive} or consistency models \cite{song2023consistency} can reduce diffusion sampling since they map numerous denoising steps in few or one direct step. Achieving such speed‐ups would facilitate real‐time deployment of these surrogates.

    \item \emph{Mixture of Experts (MoE).} Recent sparsely gated MoE architectures, exemplified by the Switch Transformer \cite{fedus2022switch}, show that a lightweight gating network can route inputs to a small subset of specialized experts, yielding massive parameter counts without proportional compute overhead. A parallel idea has already been validated for operator learning: the GNOT framework \cite{hao2023gnot} employs a gating transformer to blend multiple neural operators and thus handle multi–scale PDE regimes. Inspired by these successes, our hybrid surrogate could be extended by training several expert models—each focused on a distinct sub-regime of the parameter or solution space, while a gating network adaptively fuses their outputs. Concretely, one expert might address sub‑sonic flows, another supersonic flows, and a third transition regimes, the gate would then choose the most relevant expert or convex blend at inference time. Integrating such an MoE strategy with S‑DeepONet priors and residual video diffusion would allocate capacity where it is most needed and further generalize the framework to large‑scale, multi‑regime PDE problems.

\end{itemize}

In closing, this pioneering work highlights that combining S-DeepONet priors with diffusion‐based refinement offers a robust, flexible strategy for accurate surrogate modeling of complex, time‐dependent PDEs. By focusing on residual learning and harnessing physically meaningful conditioning, we obtain quantitatively superior predictions alongside strong fidelity, paving the way for broader adoption in accurate and robust real-time modeling and simulation, online controls, and almost instant evaluations within design and optimization loops, bypassing the high computational burden of conventional numerical methods.

\section*{Replication of results}
The data and source code supporting this study will be available in the public GitHub repository \href{https://github.com/ncsa/}{https://github.com/ncsa/} upon the paper's acceptance.

\section*{Conflict of interest}
The authors declare that they have no conflict of interest.

\section*{Acknowledgements}

Authors gratefully acknowledge the National Center for Supercomputing Applications (NCSA) at the University of Illinois for their invaluable assistance, with particular appreciation for its Research Consulting Directorate, the Industry Program, and the Center for Artificial Intelligence Innovation (CAII). Additional support was provided by the Illinois Computes project, a collaborative effort of the University of Illinois Urbana‑Champaign and the University of Illinois System. This work also leveraged both Delta and DeltaAI advanced computing and data resources, funded by the National Science Foundation (awards OAC‑2005572 and OAC‑2320345) and the State of Illinois. Delta and DeltaAI are joint initiatives of the University of Illinois Urbana‑Champaign and NCSA.

\section*{CRediT author contributions}
\textbf{Jaewan Park}: Conceptualization, Methodology, Software, Formal analysis, Investigation, Writing - Original Draft.
\textbf{Farid Ahmed}: Conceptualization, Methodology, Software, Formal analysis, Investigation, Writing - Original Draft.
\textbf{Kazuma Kobayashi}: Conceptualization, Methodology, Software, Formal analysis, Investigation, Writing - Original Draft.
\textbf{Diab Abueidda}: Conceptualization, Methodology, Supervision, Writing - Review \& Editing.
\textbf{Seid Koric}: Conceptualization, Methodology, Supervision, Resources, Writing - Original Draft, Funding Acquisition.
\textbf{Syed Alam}: Investigation, Writing - Review \& Editing.
\textbf{Iwona Jasiuk}: Supervision, Writing - Review \& Editing.

\bibliographystyle{unsrtnat}
\setlength{\bibsep}{0.0pt}
{\scriptsize \bibliography{References.bib} }

\newpage
\section*{Appendix}
\label{sec:appendix}
\appendix
\renewcommand{\thefigure}{A\arabic{figure}}
\setcounter{figure}{0}
\renewcommand{\theequation}{A\arabic{equation}}
\setcounter{equation}{0}

\subsection{Residual for each cases}

\subsubsection{ldc}
\begin{figure}[h!] 
    \centering
         \includegraphics[width=0.9\textwidth]{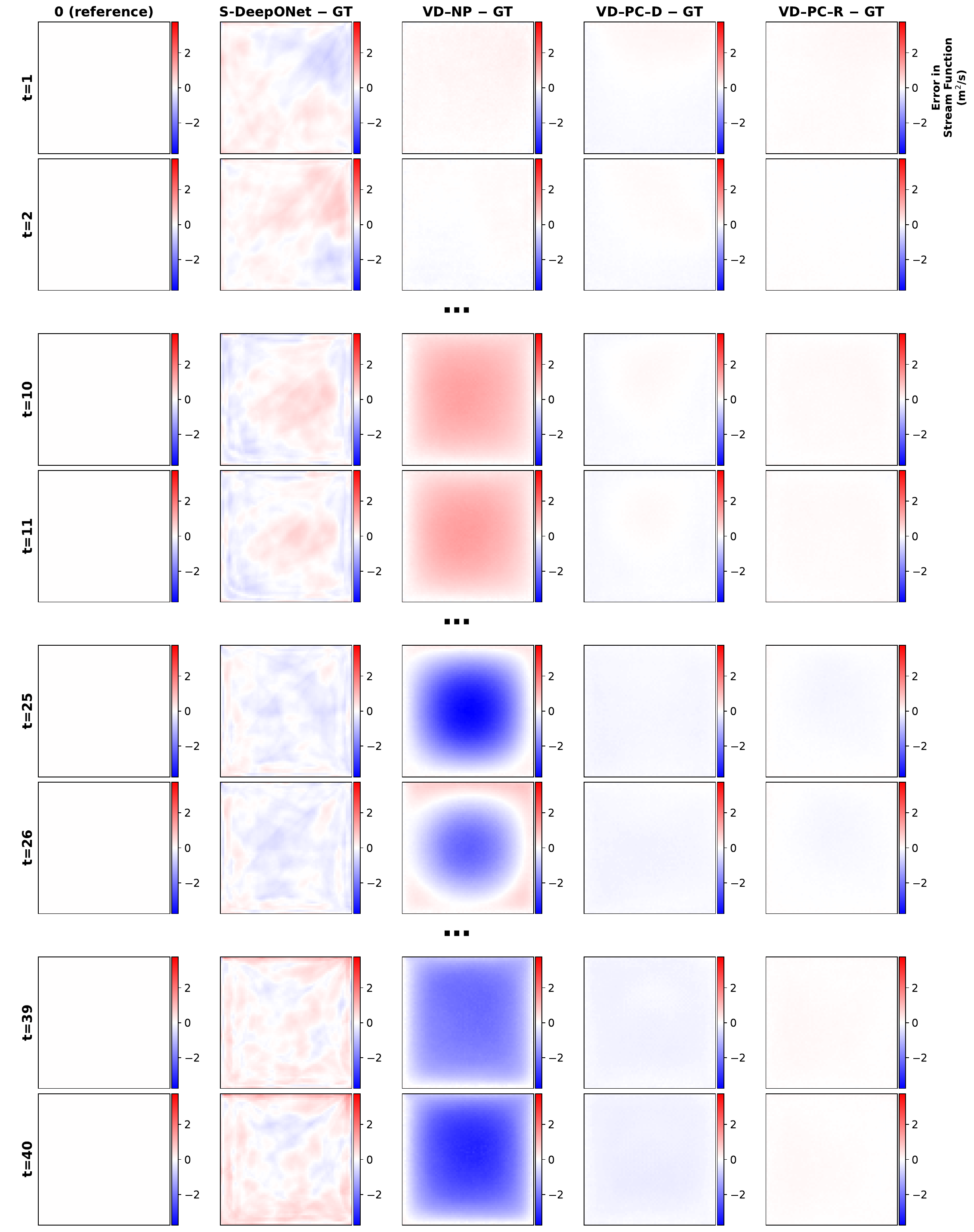}
    \caption{Qualitative comparison of stream function of best case for S-DeepONet prediction.}
    \label{appendix_ldc_quality_best}
\end{figure}

\begin{figure}[h!] 
    \centering
         \includegraphics[width=1.0\textwidth]{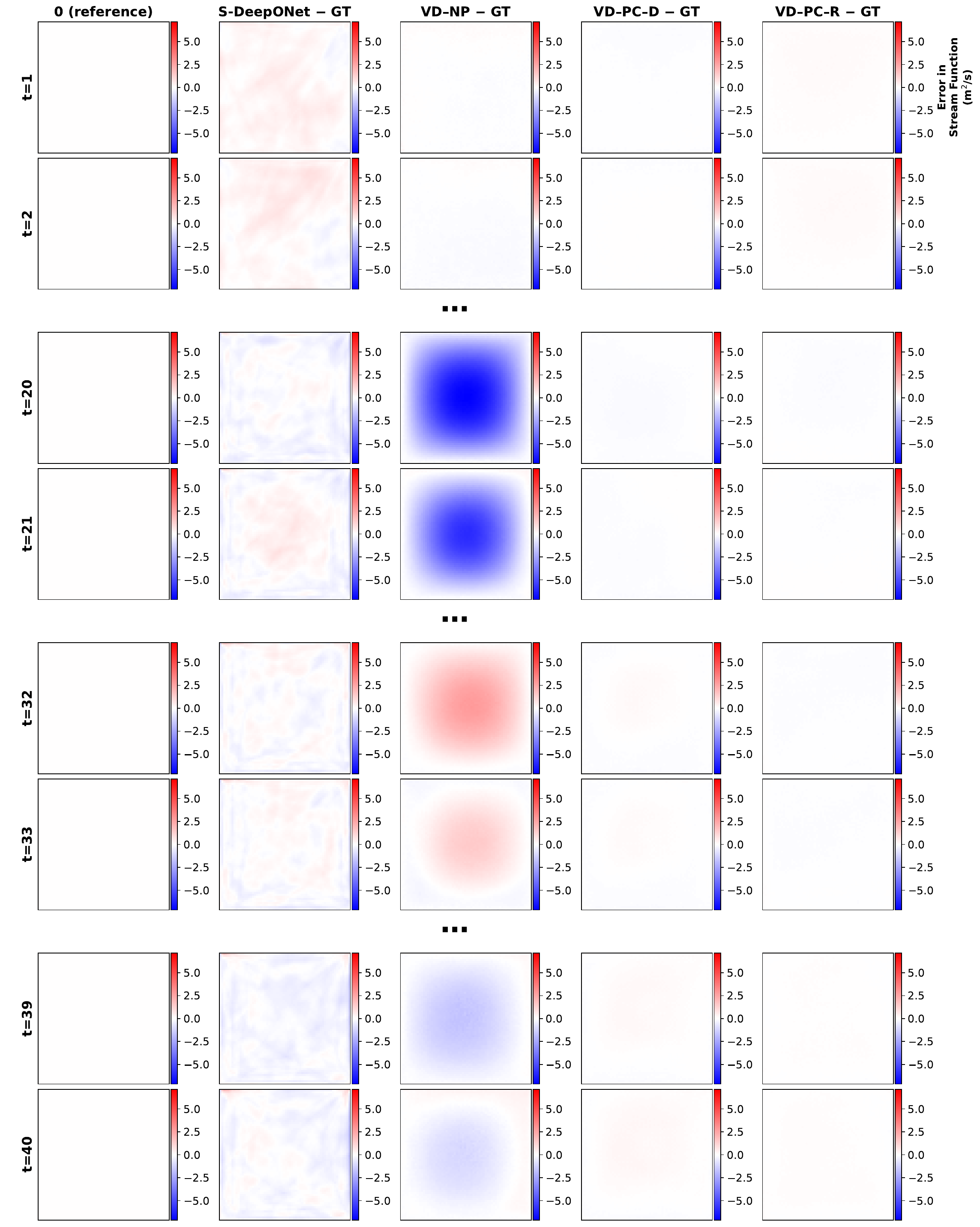}
    \caption{Qualitative comparison of stream function of median case for S-DeepONet prediction.}
    \label{appendix_ldc_quality_median}
\end{figure}

\begin{figure}[h!] 
    \centering
         \includegraphics[width=1.0\textwidth]{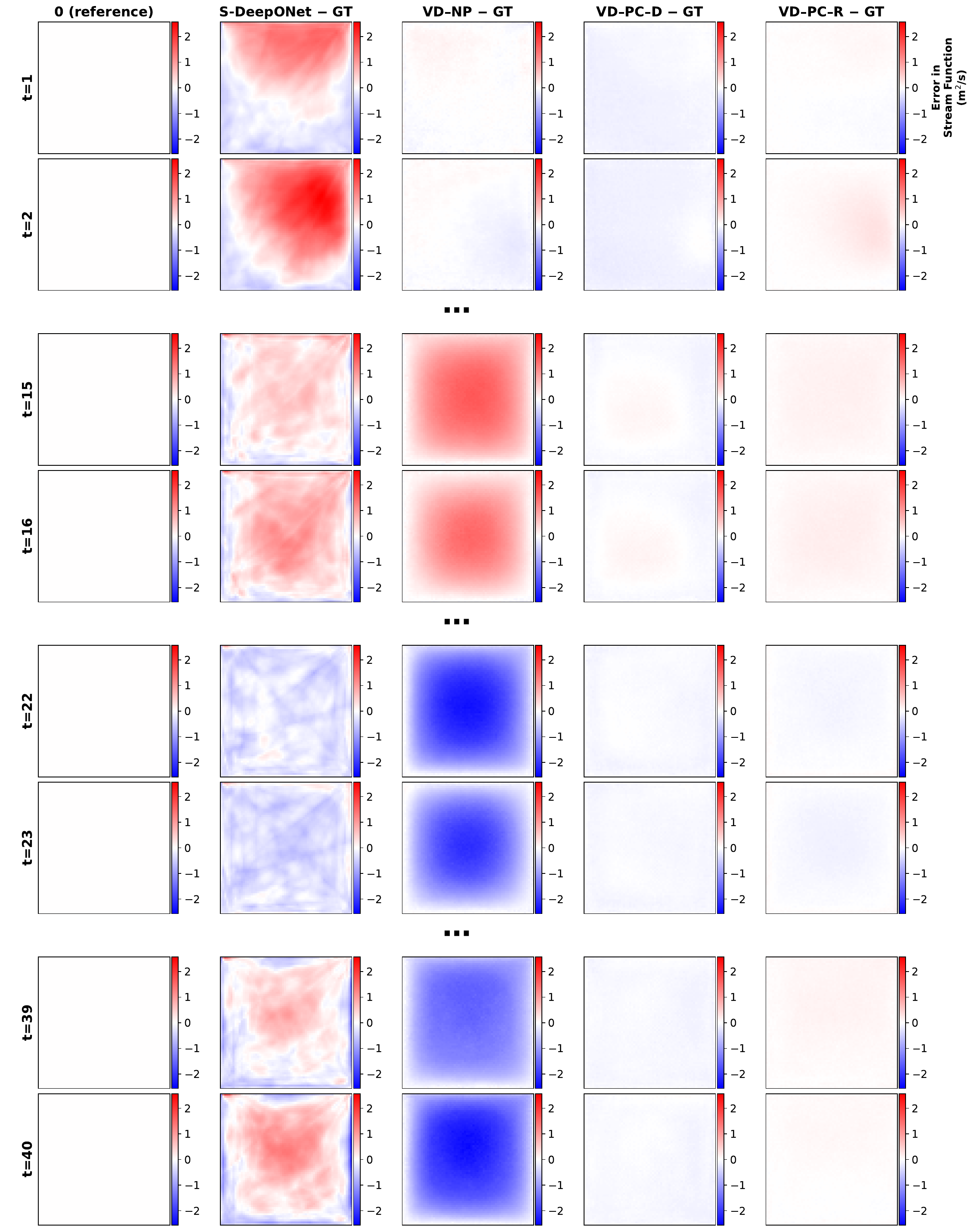}
    \caption{Qualitative comparison of stream function of worst case for S-DeepONet prediction.}
    \label{appendix_ldc_quality_worst}
\end{figure}

\subsubsection{dogbone}
\begin{figure}[h!] 
    \centering
         \includegraphics[width=1.0\textwidth]{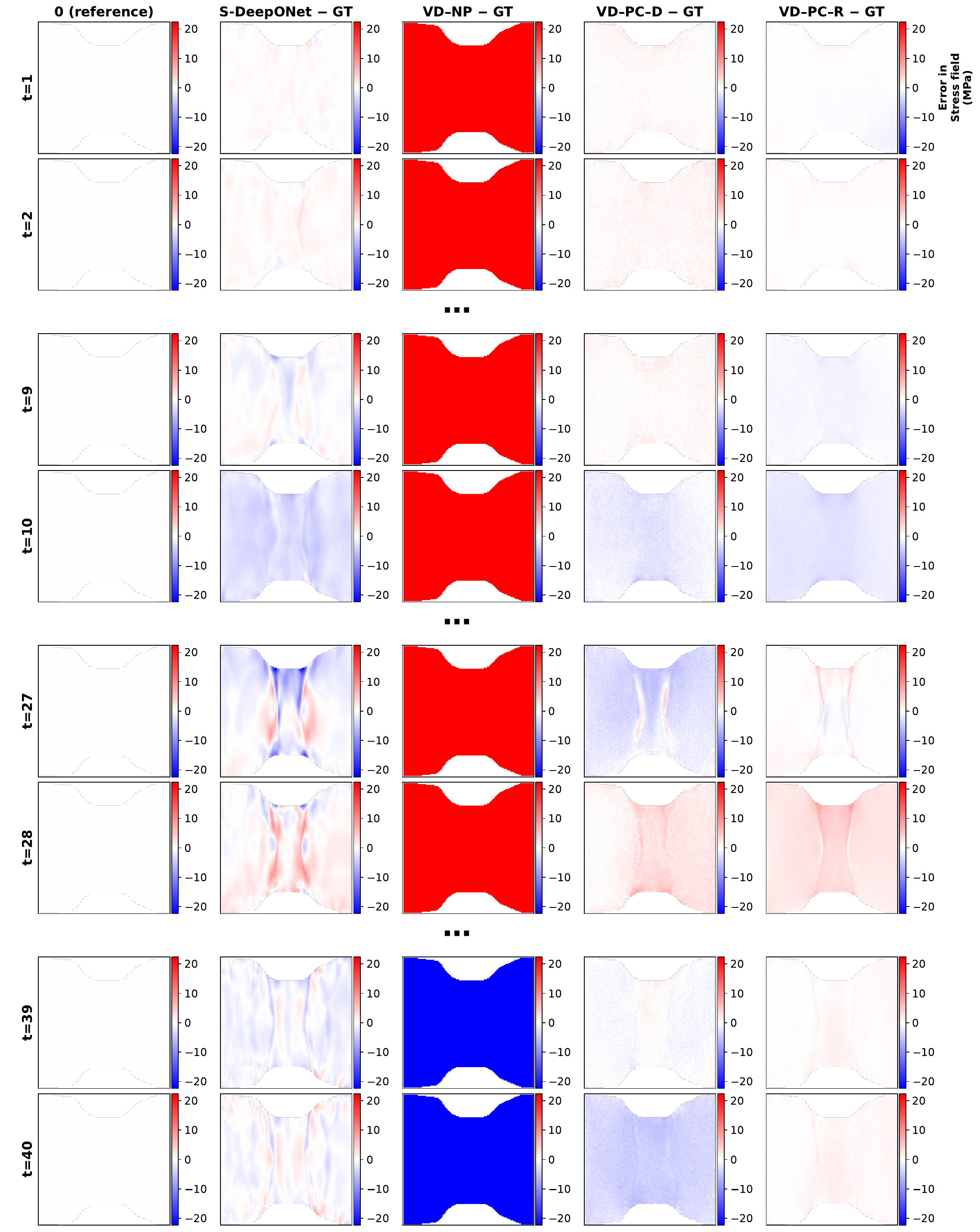}
    \caption{Qualitative comparison of von Mises stress field of best case for S-DeepONet prediction.}
    \label{appendix_dogbone_quality_best}
\end{figure}

\begin{figure}[h!] 
    \centering
         \includegraphics[width=1.0\textwidth]{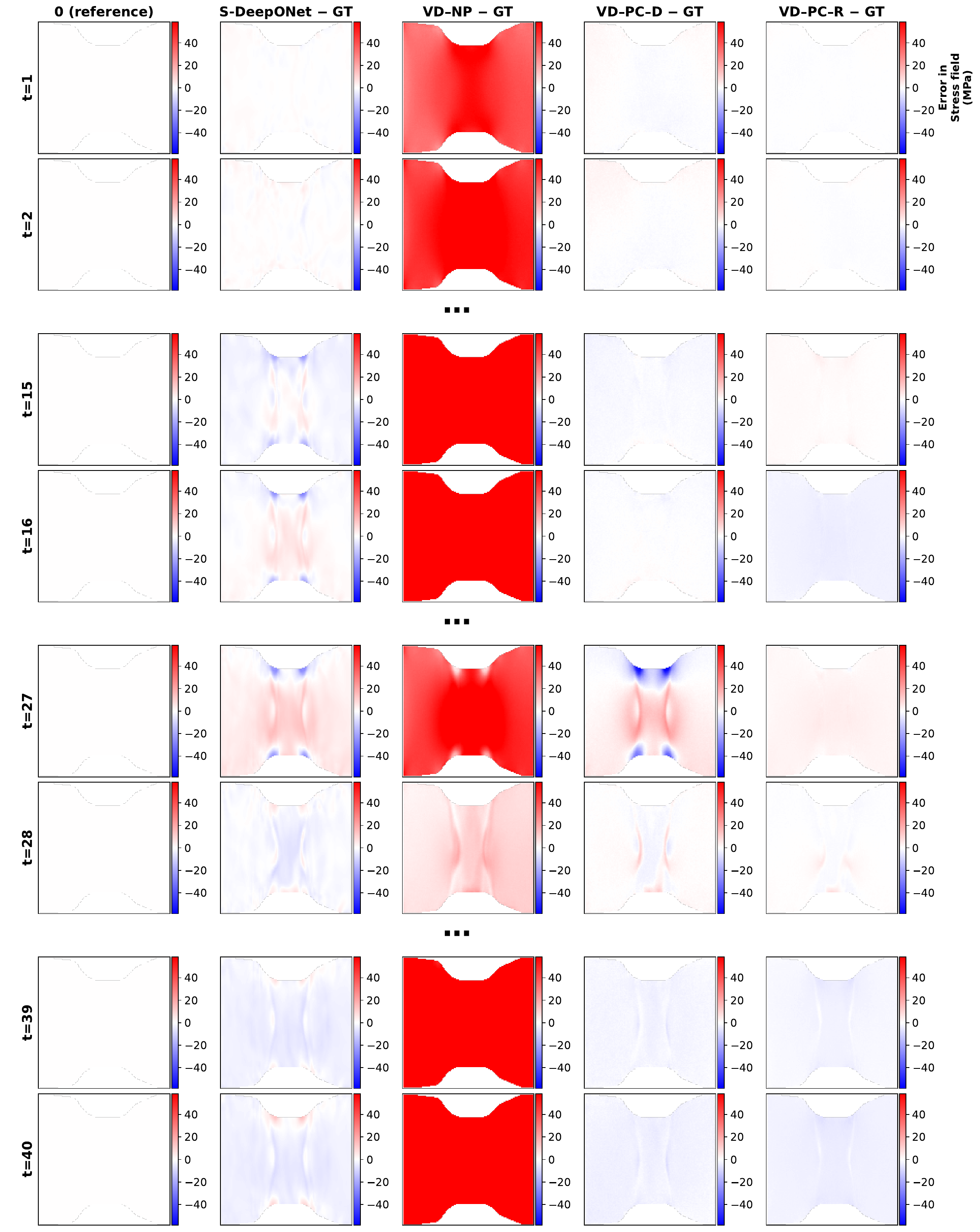}
    \caption{Qualitative comparison of von Mises stress field of median case for S-DeepONet prediction.}
    \label{appendix_dogbone_quality_median}
\end{figure}

\begin{figure}[h!] 
    \centering
         \includegraphics[width=1.0\textwidth]{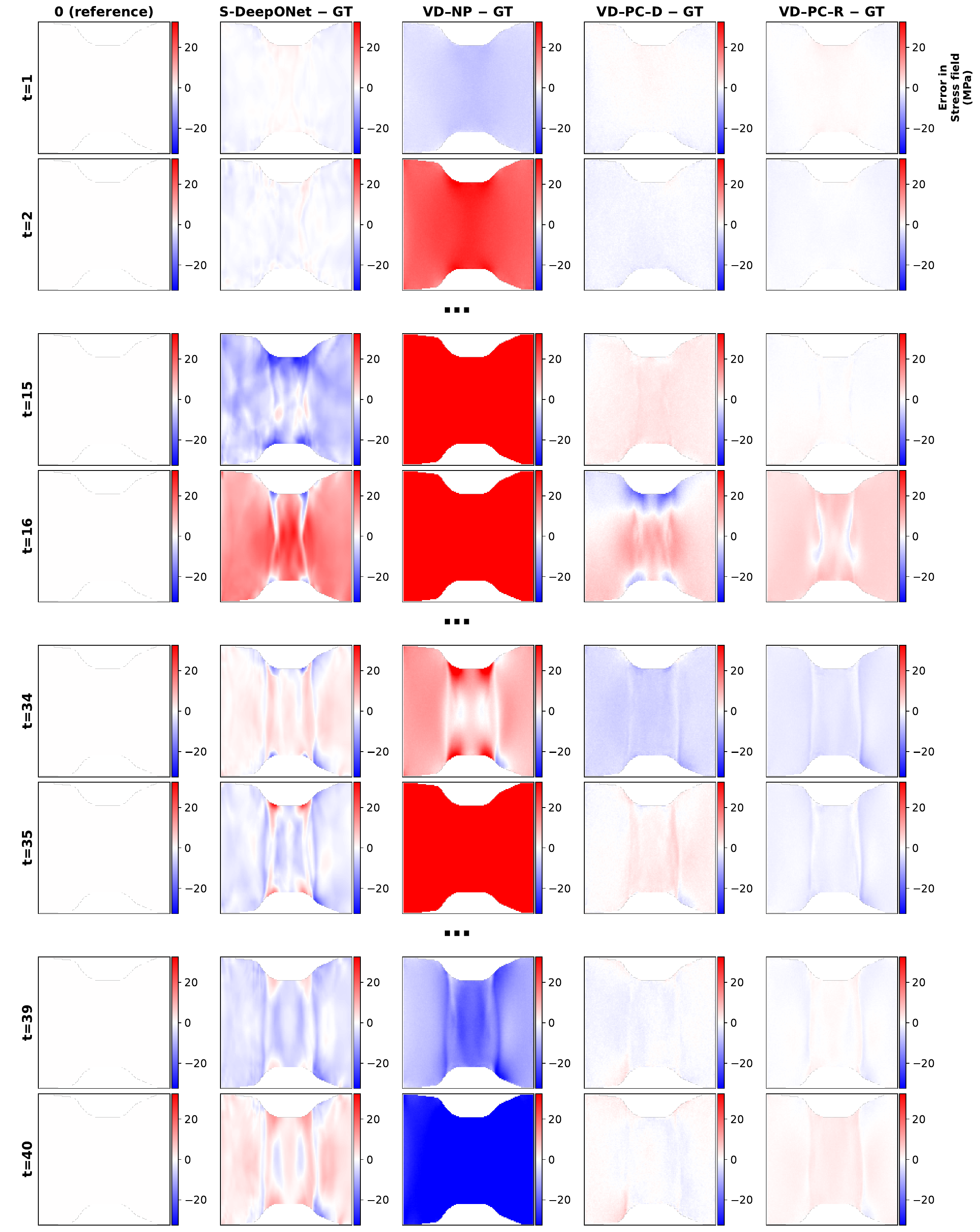}
    \caption{Qualitative comparison of von Mises stress field of worst case for S-DeepONet prediction.}
    \label{appendix_dogbone_quality_worst}
\end{figure}

\end{document}